\documentclass[mnsc,nonblindrev]{informs3}

\OneAndAHalfSpacedXI

\usepackage[plainpages=false,hyperfootnotes=false]{hyperref}
\hypersetup{
  colorlinks   = true, 
  urlcolor     = blue, 
  linkcolor    = red, 
  citecolor   = blue 
}

\usepackage{natbib}
 \bibpunct[, ]{(}{)}{,}{a}{}{,}%
 \def\bibfont{\small}%
\TheoremsNumberedThrough     
\ECRepeatTheorems

\EquationsNumberedThrough    

\MANUSCRIPTNO{}

\usepackage{dsfont}
\usepackage{amssymb}
\usepackage{amsmath}
\usepackage{enumitem}
\usepackage{booktabs}
\usepackage{siunitx}
\usepackage[colorinlistoftodos]{todonotes}
\usepackage{cleveref}

\newcommand*{\ul}{\raisebox{2pt}{\rlap{$\ulcorner$}\kern1pt}}
\newcommand*{\ur}{\raisebox{2pt}{\kern1pt \llap{$\urcorner$}}}

\newcommand*{\ulz}{\raisebox{1pt}{\rlap{$\ulcorner$}\kern2pt}}
\newcommand*{\urz}{\raisebox{1pt}{\kern1pt \llap{$\urcorner$}}}

\newcommand{\rfunc}{\overline{R}^S}
\newcommand{\brfunc}{\overline{\mathbf{R}}^S}
\newcommand{\abs}[1]{\left\lvert #1 \right\rvert}  

\newcommand{\A}{\mathsf{A}}
\newcommand{\M}{{\mathcal{M}^\infty}}

\newcommand{\F}{\mathfrak{F}}

\newcommand{\bX}{\mathbf{X}}
\newcommand{\bx}{\mathbf{x}}

\DeclareMathOperator*{\esssup}{ess\,sup}

\renewcommand{\L}{{\mathcal{L}^\infty}}
\newcommand{\im}{{\text{Im}}}

\newcommand{\bz}{{\boldsymbol{z}}}
\newcommand{\ep}{{\varepsilon}}
\newcommand{\Id}{{\mathds{1}}}

\newcommand{\E}{{\mathbb{E}}}
\newcommand{\N}{{\mathbb{N}}}
\renewcommand{\P}{{\mathbb{P}}}
\newcommand{\Q}{{\mathbb{Q}}}
\newcommand{\R}{{\mathds{R}}}

\newcommand{\x}{{\boldsymbol{x}}}

\newcommand{\z}{{\boldsymbol{z}}}

\newcommand{\X}{{\boldsymbol{X}}}

\newcommand{\Y}{{Y}}

\newcommand{\bbeta}{{\boldsymbol{\beta}}}

\newcommand{\breg}{{\overline{\bbeta}}^S}

\newcommand{\Qs}{{\mathcal{Q}}}

\newcommand{\VaR}{{\mathrm{VaR}}}
\newcommand{\ES}{{\mathrm{ES}}}

\newcommand{\supp}{{\mathrm{supp}}}

\usepackage[plainpages=false,hyperfootnotes=false]{hyperref}
\hypersetup{
  colorlinks   = true, 
  urlcolor     = blue, 
  linkcolor    = red, 
  citecolor   = blue 
}

\begin{document}

\RUNAUTHOR{Miao and Pesenti}

\RUNTITLE{Robust Elicitable Functionals}

\TITLE{Robust Elicitable Functionals}

\ARTICLEAUTHORS{%
\AUTHOR{Kathleen E.~Miao\footnote{corresponding author}}
\AFF{Department of Statistical Sciences, University of Toronto, Canada, \EMAIL{k.miao@mail.utoronto.ca}} 
\AUTHOR{Silvana M.~Pesenti}
\AFF{Department of Statistical Sciences, University of Toronto, Canada, \EMAIL{silvana.pesenti@utoronto.ca}} 
\vspace{0.5em}
February 4, 2025\footnote{First version: September 6, 2024}
} 

\ABSTRACT{
Elicitable functionals and (strictly) consistent scoring functions are of interest due to their utility of determining (uniquely) optimal forecasts, and thus the ability to effectively backtest predictions.  However, in practice, assuming that a distribution is correctly specified is too strong a belief to reliably hold.  To remediate this, we incorporate a notion of statistical robustness into the framework of elicitable functionals, meaning that our robust functional accounts for “small” misspecifications of a baseline distribution.  Specifically, we propose a robustified version of elicitable functionals by using the Kullback-Leibler divergence to quantify potential misspecifications from a baseline distribution.  We show that the robust elicitable functionals admit unique solutions lying at the boundary of the uncertainty region, and provide conditions for existence and uniqueness.  Since every elicitable functional possesses infinitely many scoring functions, we propose the class of b-homogeneous strictly consistent scoring functions, for which the robust functionals maintain desirable statistical properties. We show the applicability of the robust elicitable functional in several examples: in a reinsurance setting and in robust regression problems.
}%



\maketitle

%



\section{Introduction}

Risk measures are tools used to quantify profits and losses of financial assets.  A risk measure maps losses, stemming from, e.g. historical data or a distributional output, to a real number characterising its riskiness or in an insurance setting, e.g. the amount of assets required to be retained as reserves. However, in many situations, data or distributional information of modelled losses may be flawed for several reasons -- they may be out-of-date, sparse, or unreliable due to errors. As such, it is of interest to relax the assumption that the underlying distribution is correctly specified. In the literature, this is referred to as distributional robustness, and is often approached via a worst-case risk measure; indicatively see \cite{GhaouiOksOustry2003}, \cite{BernardPesentiVanduffel2024}, \cite{CaiLiuYin2024}. For a set of probability measures $\Qs$ -- the so-called uncertainty set -- a worst-case risk measure evaluated at a random variable $Y$ is typically defined by
\begin{equation}\label{eq:worst-case-rm}
    \sup_{\Q \in \Qs} \;R\left(F_Y^\Q\right)\,,
\end{equation}
where $R$ is a law-invariant risk measure and $F_Y^\Q$ is the cumulative distribution function (cdf) of $Y$ under $\Q$. Thus, a worst-case risk measure is the largest value the risk measure can attain over a predetermined set of cdfs.

From a statistical perspective, the notion of elicitability of a law invariant functional is of interest as it yields a natural backtesting procedure \citep{NoldeZiegel2017}. Of course, risk measures are functionals, and specific risk measures, such as the Value-at-Risk (VaR), are elicitable. An elicitable risk measure can be represented as the minimiser of the expectation of a suitable scoring function \citep{Gneiting2011}. Thus, for an elicitable risk measure $R$ with corresponding scoring function $S\colon\A\times \R \to [0, \infty)$, the worst-case risk measure  given in \eqref{eq:worst-case-rm} can equivalently be written as 
\begin{equation}\label{eq:worst-case-rm-elicitable}
        \sup_{\Q \in \Qs} \;\argmin_{z \in \A} \;\int  S(z,y)\, d F_Y^\Q(y)\,.
\end{equation}
The worst-case risk measure framework yields a worst-case cdf, that is, the one attaining the supremum in \eqref{eq:worst-case-rm} and \eqref{eq:worst-case-rm-elicitable}. This worst-case cdf may be interpreted as the distribution of the losses when the most adverse (with respect to the risk measure) probability measure in $\Qs$ materialises. Worst-case values, also called risk bounds, have been studied extensively for various choices of uncertainty set and risk measure, see \cite{RuschendorfVanduffelBernard2024} for an extensive overview.

Alternative to the worst-case methodology, we propose leveraging distributional robustness and elicitability to create a robust elicitable functional (REF). Motivated by this, we interchange the supremum and argmin in \eqref{eq:worst-case-rm-elicitable} and define the REF as
\begin{equation}\label{eq:intro-REF}
    \overline{R}^S(Y):= \argmin_{z\in \A}\;\sup_{\Q \in \Qs_\ep}  \;\int  S(z,y)\, d F_Y^\Q(y)\,,
\end{equation}
where in the exposition, we consider an uncertainty set characterised by the Kullback-Leibler (KL) divergence. In contrast to worst-case risk measures, the REF depends not only on the choice of risk measure but more specifically on the scoring function that elicits the risk measure, see also \Cref{sec: b-homog} for a detailed discussion. The optimal measure attaining the inner problem in \eqref{eq:intro-REF} can be interpreted as the, in KL divergence, largest admissible expected score; the extremal score. As the REF is the minimiser of the extremal score, it is indeed a ``robust'' risk functional. Contrary to the worst-case risk-measure, which is always larger than the baseline risk measure, the REF, which is the minimiser of the extremal score, can be smaller or larger than the baseline risk measure. From a computational perspective, if the inner optimisation problem of \eqref{eq:intro-REF} over the space of probability measures can be solved semi-analytically, the outer problem over reals can typically be tackled using classical optimisation techniques.

The benefits of elicitable functionals has been well justified in the domains of statistics and risk management. Elicitable functionals and (strictly) consistent scoring functions are of interest due to their utility of determining (uniquely) optimal forecasts, and thus the ability to effectively backtesting predictions, see e.g. \cite{Gneiting2011} and \cite{FisslerZiegel2016}.  The literature on elicitability in risk management is extensive, with many authors arguing for its importance \citep{Ziegel2016, HeKuoPeng2022}. The characterisation of elicitable convex and coherent risk measures for example has been studied in \cite{BelliniBignozzi2015}, while \cite{FisslerZiegelGneiting2016} showed that Expected Shortfall (ES) is jointly elicitable with VaR.  \cite{EmbrechtsELAL2021} explored the connection between elicitability and Bayes risks, and showed that entropic risk measures are the only risk measures that have both properties. In the related \cite{FisslerPesenti2022}, authors study sensitivity measures tailored for elicitable risk measures.
Robustness to distributional uncertainty is of concern in risk management and has been explored in the literature, see e.g. \cite{EmbrechtsPuccettiRuschendorf2013} for approximation of worst-case risk measures using the Rearrangement Algorithm, \cite{pesenti2016DM} for a discussion on distributional robustness of distortion risk measures, and \cite{EmbrechtsWangWang2015} for the aggregation robustness of VaR and ES.

Aside from its attention in DRO and the model assessment literature, the KL divergence implies that all alternative distributions have the same support as the reference distribution. This is of particular relevance in applications such as insurance settings where the loss is bounded from above and below, or in regression settings that include categorical, likert or proportional data. We refer to \Cref{subsec-reinsurance} and  \Cref{sec: rob-reg} for further details.
Distributionally robust optimisation (DRO) studies decision models where the decision maker has only partial distributional knowledge of the model at hand.  These constrained optimisation problems take the form of a minimax problem, akin to \Cref{eq:intro-REF}. A method of characterising uncertainty is to use the KL divergence, as introduced in \cite{KullbackLeibler1951}. In the realm of DRO and minimax problems, KL uncertainty is often considered, e.g. \cite{HuHong2013} study DRO problems with ambiguity sets constrained by the KL divergence, and \cite{Calafiore2007} finds mean-risk optimal robust portfolios under distributional uncertainty characterised by the KL divergence.  The minimax problem we propose in \Cref{eq:intro-REF} fits into the general framework of the DRO literature. With the REF methodology, we not only solve the inner problem semi-analytically but we also show existence and uniqueness of the outer problem.

We extend the REF to $k$-elicitable functionals, and thus obtain jointly robust VaR and ES pairs. In this setting, we obtain a probability measure that attains extremal value simultaneously for VaR and ES. This is in contrast to the classical framework of worst-case risk measures, which yields different worst-case cdfs for VaR and ES. As the proposed REF depends on the choice of scoring function, we discuss the families of homogeneous scoring function that naturally lead to visualisations of the REFs via Murphy diagrams. We showcase the applicability of the REF on a simulated reinsurance example and connect the framework to robust regression, where we explore a data driven example.
The key novelty and findings of this work is the distributionally robust elicitable functional, the REF, given in \eqref{eq:intro-REF}, which is an alternative to the classical worst-case risk measure by taking the minimiser of a worst-case expected score. We solve the inner optimisation problem of \eqref{eq:intro-REF} and obtain a semi-closed solution of the extremal probability measure. Furthermore, we prove existence and uniqueness of the REF, that is, the solution to the double optimisation problem \eqref{eq:intro-REF}. 

This paper is organised as follows; in \Cref{sec: REF}, we define the robust elicitable functional and state our main results -- solving the inner problem of \eqref{eq:intro-REF} and establishing existence and uniqueness of the REF. In \Cref{sec: b-homog}, we address the question of choosing a scoring function by considering families of parameterised scoring functions, showing that these families retain useful properties of the functional, and illustrate the behaviour of the REF in Murphy diagrams.  We extend our results to higher order elicitable risk measures such as the (VaR, ES) pair in \Cref{sec: k-elicitable}, and consider an application of the robust (VaR, ES) in reinsurance.  Finally, we explore the connection of the REF to robust regression in \Cref{sec: rob-reg}.

\section{Elicitability and robust elicitable functionals}
\label{sec: REF}
This section provides the necessary notation and recalls the notion of elicitability. We further define the robust elicitable functional and prove its existence and uniqueness.

\subsection{Robust elicitable functionals}

Let $(\Omega, \mathfrak F, \P)$ be a probability space, where we interpret $\P$ as the baseline probability measure. The baseline probability could, for example, be estimated from data or informed by expert knowledge. When not specified, expectations and distribution are taken with respect to the baseline probability $\P$. Let $\L:= \L(\Omega, \mathfrak F, \P)$ be the space of essentially bounded random variables (rvs) on $(\Omega, \mathfrak F, \P)$ and denote by $\M$ the corresponding class of cumulative distribution functions (cdfs), i.e. $\M:= \{ F ~|~ F(y) = \P(Y \le y), \; Y \in \L\}$.
 
We denote by $R \colon \M \to \A$, $\A \subseteq \R$, a potentially set-valued, law-invariant, and non-constant functional, which is a mapping from the set of admissible cdfs, $\M$, to the set of allowed predictions $\A$, also called action domain. We assume that when evaluating the functional $R(Y)$ that $Y$ is non-degenerate.
As $R$ is law-invariant, it can equivalently be defined as $R \colon \L \to \A$, by setting $R(Y):= R(F_Y)$, where $F_Y$ is the cdf of $Y \in \L$. Denote the (left-)quantile of $Y$ as $F_Y^{-1}(q) := \inf\{ y \in \R : \P (Y \leq y) \geq q\}$. We further recall the definition of the moment generating function (mgf) of a rv $Y\in \L$ evaluated at $t\in \R$ as $M_Y(t) := \E[e^{t Y}]$, and its cumulant generating function (cgf) $K_Y(t) := \log(M_Y(t))$.

Throughout we work with elicitable functionals, for this, we first recall the definition of scoring functions and elicitability. In statistics, scoring functions are used to assess the accuracy of a (statistical) estimate or prediction to the true value. Formally, $S(z,y)$ is a mapping from a prediction $z$ and a realisation $y$ of $Y$ to the non-negative reals. By convention, we view $S$ as a penalty for the forecast $z$ compared to the realised event $y$, where smaller values represent more accurate predictions.

\begin{definition}[Scoring functions and elicitability]
A scoring function is a measurable function $S \colon \A \times \R \to [0, \infty)$. A scoring function may satisfy the following properties:
\begin{enumerate}[label = $\roman*)$]
    \item A scoring function $S$ is \textbf{consistent} for a functional $R$, if 
    \begin{equation}\label{eq:scor-consistent}
     \int S(t,y) \,dF(y) \leq \int S(z, y) \,dF(y),   
    \end{equation}
    for all $F \in \M$, $t \in R(F)$, and all $z \in \A$.

    \item 
    A scoring function $S$ is \textbf{strictly consistent} for a functional $R$, if it is consistent and Equation \eqref{eq:scor-consistent} holds with equality only if $z \in R(F)$.

    \item A functional $R$ is \textbf{elicitable}, if there exists a strictly consistent scoring function $S$ for $R$.
\end{enumerate}
\end{definition}

An elicitable functional $R$ admits the representation
\begin{equation}\label{eq:armin-rep}
    R(Y) = \argmin_{z \in \A} \int S(z, y) \,dF_Y(y),
\end{equation}
for all $ F_Y \in \M$ and where $S$ is any strictly consistent scoring function for $R$. We note that $R(Y)$ can be an interval, such as in the case of the quantile, and we denote the image of $R$ by $\im R\subset \A$. In the case that $R(Y)$ is an interval, we denote the interval as $[\; R(Y)^\mathrm{l}, R(Y)^\mathrm{u}\;]$, with lower endpoint $R(Y)^\mathrm{l}$ and upper endpoint $R(Y)^\mathrm{l}$, respectively.
We call $R(Y)$ unique, if the argmin is a singleton. Many statistical functionals and risk measures are elicitable, including the mean, median, quantiles, expectiles, see e.g. \cite{Gneiting2011}, and \cite{BelliniBignozzi2015}. Thus, an elicitable risk measure is the best prediction $z$ that minimises the expected score under the baseline probability $\P$.

The definition of elicitability, in particular Equation \eqref{eq:armin-rep},  assumes knowledge of the true distribution of $Y$. However, under distributional ambiguity -- that is, uncertainty on the distribution of $Y$ -- one may wish consider a robust functional. A classical choice to model distributional deviations is the Kullback-Leibler (KL) divergence.  The KL divergence has been extensively used in model assessment; indicatively see \cite{GlassermanXu2014, Pesenti2019EJOR, BlanchetEtAl2019, LassanceVrins2023}. For a probability measure $\Q$ on $(\Omega, \F)$, the KL divergence from $\Q $ to $\P$ is defined as
\begin{equation*}
    D_{KL}(\Q ~||~ \P) := \E^\P\left[\frac{d\Q}{d\P} \log\left(\frac{d\Q}{d\P}\right)\right]\,,
\end{equation*}
if $\Q$ is absolutely continuous with respect to $\P$ and $+\infty$ otherwise. We denote by $\E^\Q[\cdot]$ the expected value under $\Q$, and for simplicity write $\E[\cdot] := \E^\P[\cdot]$, when considering the baseline probability measure $\P$. We use the KL divergence to describe an uncertainty set and propose the following robust elicitable functionals. 

\begin{definition}[Robust elicitable functional]
 Let $R$ be an elicitable functional with strictly consistent scoring function $S \colon \A \times \R \to [0, \infty)$ and let $\ep\ge 0$. Then we define the \textbf{robust elicitable functional} (REF), for $Y \in \L$, by
\begin{equation}\label{opt:KL}
    \rfunc(Y) := \argmin_{z \in \A}\; \sup_{\Q \in \Qs_\ep}\; \E^\Q\left[S(z, Y)\right]\,,
\end{equation}
where the uncertainty set $\Qs_\ep$ is given as
\begin{equation}
\label{eq:uncertainty-set}
\Qs_\ep := \big\{\Q ~|~\Q \ll \P,\quad \text{and} \quad  D_{KL}(\Q ~||~ \P) \le \ep \big\}\,.
\end{equation}
\end{definition}

The uncertainty set $\Qs_\ep$ contains all probabilities that have a KL divergence to $\P$ less than or equal to $\ep$. For simplicity, we omit the dependence of $\Qs_{\ep}$ on $\P$ as the baseline is fixed throughout. The parameter $\ep \ge0$ is a tolerance distance that quantifies the distance to the baseline probability $\P$. Clearly, if $\ep = 0$, only the baseline measure is considered and we recover the traditional definition of elicitable functionals. The uncertainty set $\Qs_\ep$ is compact under the topology of weak convergence, for all $\ep \in (0, \infty)$, see \cite{vanErvenHarremoes2012}. From \eqref{opt:KL}, we see that the REF is the best prediction $z$, that minimises the expected score if the universe chose the most adverse distribution of $Y$. Thus, the $\rfunc(Y)$ may be larger or smaller than $R(Y)$, hence it is not a worst-case functional in the sense of \eqref{eq:worst-case-rm-elicitable}, which is always larger than $R(Y)$.

When $Y$ has a positive probability mass at its essential supremum, then for a too large tolerance distance, as specified in the next proposition, the inner problem in \eqref{opt:KL} becomes degenerate. The next two results make this precise.
\begin{lemma}\label{lemma:limit-KL}
  Let $Y$ be a rv satisfying $\P(Y = \esssup Y)  =: p>0 $ and define the probability measure
  \begin{equation*}
      \frac{d \Q^s}{d \P} := \frac{e^{s Y}}{\E[e^{s Y}]}\,. 
  \end{equation*}
Then the following limits hold
  \begin{equation*}
      \lim_{s \to 0} D_{KL}(\Q^s ~||~ \P) = 0\,
      \quad \text{and} \quad
            \lim_{s \to +\infty} D_{KL}(\Q^s ~||~ \P) = \log\big(\tfrac{1}{p}\big)\,.
  \end{equation*}
\end{lemma}

\textit{Proof}
The fact that $\lim_{s \to 0} D_{KL}(\Q^s ~||~ \P) = 0$ follows as $\lim_{s \to 0}\frac{d \Q^s}{d \P} = 1$. For the second limit, we redefine the rv $Y$ via
\begin{equation*}
    Y(\omega) = \begin{cases}
        W(\omega) & \text{ on } \omega \in A^\complement \\
        \bar{y} := \esssup Y & \text{ on } \omega \in A\,,
    \end{cases}
\end{equation*}
where $W:= Y\Id_{\{Y < \esssup Y\}}$, $\P(Y \in A) = p$, and $A^\complement$ denotes the complement of $A$. By setting $W = 0$ on $A$, we have that $\E[e^{sW}] =: M_W(s)$ is the mgf of $W$. For $\omega \in A^\complement$, define $v:= W(\omega)$ then we obtain 
\begin{align*}
    \frac{d \Q^s}{d \P}(\omega) &= \frac{e^{s v}}{M_W(s) + p e^{s\bar{y}}}
        = \frac{1}{M_{W-v}(s) + p e^{s(\bar{y}-v)}}\;.
\end{align*}
As $M_{W-v}(s) \geq 0$ and $\bar{y} > v $, we have that $\lim_{s \to +\infty} \frac{d \Q^s}{d \P} = 0 $ $\P$-a.s. on $ A^\complement$.
For $\omega \in A$, we  have
\begin{align*}
    \frac{d \Q^s}{d \P} (\omega) &= \frac{e^{s\bar{y}}}{M_W(s) + p e^{s\bar{y}}}
    = \frac{1}{M_{W-\bar{y}}(s)  + p}\,.
\end{align*}
Since $\bar{y} > W$ $\P$-a.s., it holds that
\begin{equation*}
    \lim_{s \to +\infty}M_{W-\bar{y}}(s)=
    \E\big[\lim_{s \to +\infty} e^{s(W-\bar{y})} \big]
    = 0 
    \;, 
\end{equation*}
and $\lim_{s \to +\infty} \frac{d \Q^s}{d \P} = \frac{1}{p} $ $\P$-a.s. on $A$.    
Thus, we conclude that $\lim_{s \to +\infty} D_{KL}(\Q^s ~||~ \P) = \log(\frac{1}{p})$.
\hfill \Halmos

This lemma shows that if $Y$ takes positive probability mass at its essential supremum, the KL-divergence converges to $\log(\frac{1}{p})$, which provides an upper limit on the choice of tolerance distance $\ep$, as detailed in the next assumption. We note that if $Y$ does not have a positive probability mass at its essential supremum, then there are no restrictions on the tolerance distance.

\begin{assumption}[Maximal Kullback-Leibler distance]\label{asm}
Define $p(z) := \P\big(S(z,Y) = \esssup S(z,Y)\big)$. Then one of the following holds:
\begin{enumerate}[label = $\roman*)$]
    \item    If $p(z) = 0$ for all $z \in \A$, then the tolerance distance satisfies $\ep \in[0, \infty)$. 

    \item If $p(z) > 0$ for some $z \in \A$, then the tolerance distance $\ep$ satisfies 
    $$0\le \ep <  \log \left(\tfrac{1}{p(z)}\right)\,
    \quad \text{for all} \quad z \in \A\,.
    $$
\end{enumerate}
\end{assumption}

\Cref{lemma:limit-KL} and Assumption \ref{asm} give rise to an interpretation of the choice of $\ep$, which depends on the baseline distribution of $S(z,Y)$. That is, when $\ep = 0$, one is completely confident in the baseline probability, while under case $ii)$ of Assumption \ref{asm}, $\ep = \log\left(\frac{1}{p(z)}\right)$ corresponds to maximal uncertainty. Note that if $p(z) = 0$, then there is no upper limit on the tolerance distance and $\ep \in[0,\infty)$, with maximal uncertainty corresponding to $\ep = \infty$.

The next result shows that for $Y$ with positive probability mass at its essential supremum, if the tolerance distance of the uncertainty set is too large, i.e. $\ep \ge \log(\frac{1}{p(z)})$, then the worst-case probability measure puts all probability mass on the essential supremum of $Y$. Thus if $\ep \ge \log(\frac{1}{p(z)})$, the REF defined in \eqref{opt:KL} is degenerate justifying Assumption \ref{asm} $ii)$.
\begin{proposition}
Let $p(z)>0 $ and be given as in Assumption \ref{asm}.
If $\ep \ge \log\big(\frac{1}{p(z)}\big)$, then 
    the optimal probability measure $\Q^\dagger$ attaining the inner supremum in \eqref{opt:KL} has Radon-Nikodym density
    \begin{equation*}
        \frac{d\Q^\dagger}{d\P} = \frac{\Id_{\{S(z,Y) = \esssup S(z,Y)\}}}{p(z)}\,.
    \end{equation*}

\end{proposition}

\textit{Proof}
First, $\Q^\dagger$ lies within the KL uncertainty set, since
\begin{align*}
 D_{KL}(\Q^\dagger ~||~ \P)  
    &=
    \E\left[\frac{\Id_{\{S(z,Y) = \esssup S(z,Y)\}}}{p(z)} \log\left(\frac{\Id_{\{S(z,Y) = \esssup S(z,Y)\}}}{p(z)}\right)\right]
    =\log\left(\tfrac{1}{p(z)}\right)   \,.
\end{align*}
Next, $\E^{\Q^\dagger}[S(z,Y)] = \esssup S(z,Y)$ 
and moreover, for any probability measure $\Q$ that is absolutely continuous with respect to $\P$, it holds that $\E^\Q[S(z,Y) ]\le \esssup S(z,Y)$. Hence, $\Q^\dagger$ attains the inner supremum in \eqref{opt:KL}.
\hfill\Halmos

In the case when the tolerance distance satisfies Assumption \ref{asm}, the probability measure attaining the inner supremum in \eqref{opt:KL} becomes non-degenerate and the REF admits an alternative representation discussed next.

\begin{theorem}[Kullback-Leibler Uncertainty]
\label{thm:KLUncertainty}
Let $S\colon \A \times \R \to [0, \infty)$ be a strictly consistent scoring function for $R$, and $\ep$ such that Assumption \ref{asm} is satisfied. 
Then, the REF has representation
\begin{equation*}
    \rfunc(Y) = 
    \argmin_{z \in \A}\; \frac{ \E\left[S(z, Y) e^{\eta^*(z)S(z, Y) }\right]}{\E\left[e^{\eta^*(z)S(z, Y) }\right]}
    \,,
\end{equation*}
where for each $z\in \A$, $\eta^*(z) \ge 0$ is the unique solution to $\ep = D_{KL}(\Q^{\eta(z)}~||~\P)$, with
\begin{equation*}
    \frac{d\Q^{\eta(z)}}{d\P}
    :=
    \frac{ e^{\eta(z) S(z, Y) }}{\E\left[e^{\eta(z) S(z, Y)  }\right]}\,.
\end{equation*}
\end{theorem}

\textit{Proof}
Let $Y $ have cdf $F$ under $\P$ and denote its (not-necessarily continuous) probability density function (pdf) by $f$, such that e.g. $\E[Y] = \int y f(y)\, \nu(dy)$, where integration is tacitly assumed to be over the support of $Y$, which we denote by $\supp(Y)$, and where $\nu$ is either the counting or the Lebesgue measure, depending whether $Y$ is a continuous or discrete rv. Then, the inner problem of \eqref{opt:KL} can be written as an optimisation problem over densities as follows
\begin{align*}
    \sup_{g\colon \R \to \R \; }\; \int S(z, y) g(y)\nu(dy)\,,
    \quad \text{subject to}\quad
    &\int \frac{g(y)}{f(y)}\log\left( \frac{g(y)}{f(y)}\right) f(y) \, \nu(dy) \le \ep\,, 
    \\[0.5em]
    &\int g(y) \nu (dy) = 1\,, 
    \quad \text{and}
    \\[0.5em]
    &g(y) \ge 0\,, \quad \text{for all } y \in \supp(Y)\,.
\end{align*}
The above optimisation problem admits the Lagrangian with Lagrange parameters $\eta_1, \eta_2 \ge 0$, and $\eta_3(y)\ge0$, for all $y \in \supp(Y)$, as
\begin{equation*}
    L(\eta_1, \eta_2, \eta_3, g) = 
    \int \left(- S(z, y) g(y)
     + \eta_1\log\left(\frac{g(y)}{f(y)}\right) g(y) 
     +\eta_2 g(y) 
    - \eta_3(y) g(y)\right)
    \;\nu(dy)
    - \eta_1 \ep - \eta_2\,.
\end{equation*}

The Lagrange parameter $\eta_3(y)$ guarantees, for all $y \in \supp(Y)$, the non-negativity of $g(\cdot)$, that is $g(y) \ge 0$, whenever $f(y) > 0$, and $g(y) = 0$ otherwise. The associated Euler-Lagrange equation  becomes
\begin{align*}
    - S(z, y) 
     + \eta_1\left(\log\left(\frac{g(y)}{f(y)}\right)  + 1\right)
     +\eta_2 
    + \eta_3(y)  = 0\,.
\end{align*}

This implies that
\begin{equation*}
\frac{g(y)}{f(y)}
    =
    \exp\left\{\frac{1}{\eta_1}\left(S(z, y) - \eta_2 - \eta_3(y)\right) - 1\right\}\,.
\end{equation*}
Thus, $\eta_3(y) \equiv 0$ and imposing $\eta_2$ yields
\begin{equation*}
\frac{g(y)}{f(y)}
    =
    \frac{ \exp\left\{\frac{1}{\eta_1}S(z, y) - 1\right\}}{\int\exp\left\{\frac{1}{\eta_1}S(z, y)  - 1\right\} f(y) \nu(dy)}\,. 
\end{equation*}

Reparametrising $\eta := \frac{1}{\eta_1}$, we define the Radon-Nikodym derivative
\begin{equation*}
    \frac{d\Q^{\eta}}{d\P}
    :=
    \frac{g(Y)}{f(Y)}
    =
    \frac{ e^{\eta S(z, Y) }}{\E\left[e^{\eta S(z, Y)  }\right]}\, ,
\end{equation*}
which implies that $Y$ under $\Q^{\eta}$ has density $g$.

Next, we show that for each $z \in \A$, the optimal Lagrangian parameter $\eta^*$ is the solution to $\ep = D_{KL}(\Q^{\eta} ~||~ \P)$, i.e. that KL divergence constraint is binding. For this we first show that for fixed $z \in \A$, $D_{KL}(\Q^{\eta} ~||~\P) $ is strictly increasing in $\eta$. We calculate
\begin{align}
    D_{KL}(\Q^\eta ~||~\P) &= \E\left[\frac{ e^{\eta S(z, Y) }}{\E\left[e^{\eta S(z, Y) }\right] } \log\left(\frac{ e^{\eta S(z, Y) }}{\E\left[e^{\eta S(z, Y) }\right] }\right)\right]
    \nonumber
    \\
    &=
    \E\left[\frac{e^{\eta S(z,Y)}}{\E[e^{\eta S(z,Y)}]}\Big(\eta S(z,Y) - \log\left(\E[e^{\eta S(z,Y)}]\right)\Big)\right]
    \nonumber
    \\
    &= 
    \eta \frac{\E[e^{\eta S(z,Y)} S(z,Y)]}{\E[e^{\eta S(z,Y)}]} - \log\left(\E[e^{\eta S(z,Y)}]\right)
    \nonumber
    \\
    &= \eta  K'_{S(z,Y)}(\eta) - K_{S(z,Y)}(\eta) =: d(\eta) \label{eq:d}  \,, 
\end{align}
where $K_{S(z,Y)}(\eta) := \log\left(\E[e^{\eta S(z,Y)}]\right)$, and $K'_{S(z,Y)}(\eta) := \frac{\partial}{\partial \eta} K_{S(z,Y)}(\eta)$ denotes the derivative with respect to $\eta$. The interchange of the differential operator and expectation is valid as we assume that $Y$ is essentially bounded. Observe that $K_{S(z,Y)}(\cdot)$ is the cgf of $S(z,Y)$, thus it is differentiable and strictly convex. Therefore $d(\eta)$ is increasing since 
\begin{align*}
    d'(\eta) &= K'_{S(z,Y)}(\eta) + \eta K''_{S(z,Y)}(\eta) - K'_{S(z,Y)}(\eta)= \eta K''_{S(z,Y)}(\eta) > 0.
\end{align*}
Furthermore, as the objective function of the inner problem equals the derivative of the cgf, it is also increasing in $\eta$. Indeed, 
\begin{align*}
\frac{\E\left[S(z, Y) e^{\eta S(z, Y) }\right]}{\E\left[e^{\eta S(z, Y) }\right]}= K'_{S(z,Y)}(\eta)\,,
\end{align*}
which is increasing in $\eta$. Thus, both the objective function of the inner problem and the KL divergence are strictly increasing in $\eta$, the constraint is binding and $\eta^*$ is the unique solution to $\ep = D_{KL}(\Q^{\eta}~||~\P)$. To see that a solution to $\ep = D_{KL}(\Q^{\eta}~||~\P)$ exists, note that by \Cref{lemma:limit-KL}, it holds that $\lim_{\eta \to 0} D_{KL}(\Q^\eta || \P) = 0$ and  $\lim_{\eta \to +\infty} D_{KL}(\Q^\eta || \P) = \log(\frac{1}{p(z)})$.

As $\eta$ and $\Q^{\eta}$ depends on $z$, we make this dependence explicit in the statement, and write $\eta(z)$ and $\Q^{\eta(z)}$.
\hfill\Halmos

\begin{remark}
\Cref{thm:KLUncertainty} assumes that $Y\in \L$, though this is an assumption that can be relaxed. Indeed, from the proof of \Cref{thm:KLUncertainty}, we see that we only require the mgf of $S(t,Y)$ at $\eta^*(t)$, i.e., $\E[e^{\eta^*(t)S(t,Y)}]$, to be finite for all $t$ in the neighbourhood of the optimal $z^*$. 
Moreover, there are choice of  pairs $(S, Y)$ of scoring functions and rv for which the REF exists and where $Y$ is not essentially bounded. 
An example is $Y\sim N(\mu, \sigma)$ normally distributed with mean $\mu$ and standard deviation $\sigma$, and the pinball loss, $S(z,y) = \big(\Id_{\{y \le z\}} - \alpha\big) (z-y)$, $\alpha \in (0,1)$.  Clearly, $Y$ is not essentially bounded, however, the mgf of the scoring function is finite. To see this, for $t\in \R$, we have $\E[e^{t{S(z,Y)}}] = \E[e^{t{(1-\alpha)(z-Y)^+ + t\alpha(z-Y)^-}}] \leq \E[e^{t (1-\alpha)(z-Y) + t\alpha(z-Y)}] = \E[e^{t (z-Y)}] = e^{tz} M_Y(-t)$ where $(\cdot)^+$ and $(\cdot)^-$ denote the positive and negative part respectively, and $M_Y(\cdot)$ the mgf of $Y$. As the mgf of $Y$ equals $M_Y(t) = e^{t\mu + \frac{1}{2}\sigma^2 t^2}$, is finite for all $t \in\R$ and \Cref{thm:KLUncertainty} applies.
\end{remark}

\subsection{Existence and uniqueness}
For existence and uniqueness of the REF we require additional properties on the scoring functions. Thus, we first recall several definitions related to scoring functions. The first set of properties rely on the first order condition of elicitable functionals. Indeed, as elicitable functionals are defined via an argmin, they can often be found by solving the corresponding first order condition. The following definition of identification functions make this precise. We refer the interested reader to \cite{SteinwartEtAl2014} for details and discussions.

\begin{definition}[Identification function]
Let $R$ be an elicitable functional.  Then, a measurable function $V: \A \times \R \to \R$ is

\begin{enumerate}[label = $\roman*)$]

 \item  called an \textbf{identification function} for $R$, if 
 \[
    \E[V(z, Y)] = 0 \;\text{ if and only if } \; z \in R(Y)\,,
 \]
for all $z \in \im^\circ R$, where $\im^\circ R$ is the interior of $\im R$, and for all $Y \in \L$.

\item \textbf{orientated} if $V$ is an identification function and
\[
\E[V(z, Y)] > 0 \;\text{ if and only if } \;z > R(Y)^u\,,
\]
for all $z \in \im^\circ R$, and for all $Y \in \L$. 

\item The functional $R$ is called 
 \textbf{identifiable} if there exists an identification function for $R$.
\end{enumerate}
\end{definition}

An identification function thus characterises the first order condition of elicitable functionals and is therefore intimately connected to scoring functions. Moreover, an oriented identification function can further rank predictions. To see this, note that any identification function gives raise to a scoring function, and in particular, any oriented identification function gives raise to an order sensitive scoring function, see e.g. \cite{SteinwartEtAl2014}. A scoring function is called \textbf{order sensitive} (or accuracy rewarding) for $R$, if for all $t_1, t_2 \in \A$ with either $R(Y)^\mathrm{u} < t_1<t_2$ or $R(Y)^\mathrm{l}> t_1>t_2$ it holds that 
\begin{equation*}
    \E\big[S\big(R(Y), Y\big)\big]< \E\big[S\big(t_1, Y\big)\big]< \E\big[S\big(t_2, Y\big)\big]\,.
\end{equation*}
Thus, the further away the prediction $t_2$ is from $R(Y)$, the larger its expected score.

Moreover, recall that a scoring function S is locally Lipschitz continuous in $z$, if for all intervals $[a,b] \subseteq \A$, there exists a constant $c_{a,b}\ge0$ such that for all $z_1, z_2 \in [a,b]$ and all $y \in \supp(Y)$, we have
\begin{equation*}
    \big|S(z_1, y) - S(z_2, y)\big| \le c_{a,b} \, |z_1 - z_2|\,.
\end{equation*}

We next establish the following result, which characterises identification functions of exponentially transformed scoring functions, that will be instrumental in proving existence of the REF.

\begin{lemma}
\label{lemma-indentification}
 Let $S: \A \times \R \to [0, \infty)$ be strictly convex in and locally Lipschitz continuous in its first component, and a strictly consistent scoring function for a functional $R$.
 Moreover, define $H(z,y):= e^{v\, S(z,y)}$, for $v>0$.

Then, $H(\cdot, \cdot)$ is strictly convex in $z$ and locally Lipschitz continuous in $z$. Moreover, $R^H:= \argmin_{z\in A} \E[H(z,Y)]$ exists for $\A \subseteq \R$, thus $R^H$ is elicitable with scoring function $H(\cdot, \cdot)$. Furthermore, there  exists an oriented identification function $W$ for $R^H$ that satisfies
\begin{equation*}
    k(z) \, W(z,y) = v e^{v\, S(z,y)}\,  \frac{\partial}{\partial z} S(z,y)\,, \quad \text{for almost all }\quad (z,y) \in \im^\circ R \times \supp(Y)\, ,
\end{equation*}
and for some $k(z)> 0$.
\end{lemma}

\textit{Proof}
Clearly, $H$ is strictly convex in $z$ as it is a composition of the monotone and convex exponential function and the scoring function, which is convex in $z$.  Similarly, it is locally Lipschitz continuous in $z$ as $H$ is the composition of two locally Lipschitz functions.  As $H$ is strictly convex in $z$, we have that $\E[H(z,Y)]$ is also convex in $z$ for all $Y \in \L$. Thus, $R^H$ exists and is an elicitable functional. 

By Corollary 9 of \cite{SteinwartEtAl2014}, $R^H$ is identifiable and has an oriented identification function, which we denote by $W$.  By iii), Theorem 8 of \cite{SteinwartEtAl2014}, the identification function satisfies for almost all $(z,y)$
\[
k(z)\,W(z,y) =  \frac{\partial}{\partial z} H(z,y) = v  e^{v S(z,y)} \frac{\partial}{\partial z}S(z,y)\,, 
\]
for some $k(z) > 0$, $z\in \A$, and where the last equation holds by definition of $H$. 
\hfill \Halmos

The next results shows the conditions for the REF to exist, and when it is unique.

\begin{theorem}[Existence and uniqueness]
\label{thm:uniqueness}
Let $R$ be an elicitable functional with strictly consistent scoring function $S$, and $\ep$ such that Assumption \ref{asm} is satisfied. Further assume that $S(z,y)$ is strictly convex in $z$, and continuously differentiable in $z$.
Assume that
 $\int_\A |\frac{\partial}{\partial z }S(z,y)| dz <\infty$, for all $y\in \supp(Y)$. 
 Then the following hold:
\begin{enumerate}[label = $\roman*)$]
    \item there exists a solution to optimisation problem \eqref{opt:KL}, that is $\rfunc(Y)$ exists,

    \item if $\argmin_{z \in \A} \E[ e^{v\, S(z,Y)}]$ is a singleton for all $v >0$, then the solution to optimisation problem \eqref{opt:KL} is unique.
\end{enumerate}
\end{theorem}

\textit{Proof}
We define the value function of the inner optimisation problem of \eqref{opt:KL} by 
\begin{equation}\label{eq:J}
   J(z):=  \sup_{\Q \in \Qs_{\ep}} \E^\Q[S(z,Y)] \,.
\end{equation}
For existence $i)$ we first apply the envelope theorem for saddle point problems, Theorem 4 in \cite{MilgromSegal2002} to derive an expression for $\frac{d}{dz} J(z)$ and second show that $\frac{d}{dz} J(z)$ crosses zero. For $ii)$ we show that $\frac{d}{dz} J(z)$ crosses zero at most once, thus $J(z)$ admits a unique minima.

\underline{Part 1:}
Rewriting optimisation problem \eqref{eq:J} as a constrained optimisation problem over densities gives
\begin{equation*}
    \sup_{g \; \text{density}} \int S(z,y) g(y) \nu(dy)
    \,, \quad \text{subject to} \quad
    \int g(y) \log\left(\frac{g(y)}{f(y)}\right) \nu(dy ) \le \ep
    \,.
\end{equation*}
 As the space of densities is convex and as the objective function $\mathcal{J}(g,z) := \int S(z,y) g(y) \nu(dy)$ and the constraint function $c(g,z) := \ep - \int_\R g(y) \log\left(\frac{g(y)}{f(y)}\right) dy$ are both concave in $g$, the constrained optimisation problem can be represented as a saddle-point problem with associated Lagrangian
 \begin{equation*}
     L(g,\eta, z) := \mathcal{J}(g,z) + \eta c(g,z)\,.
 \end{equation*}
Moreover, it holds that
\begin{equation*}
    J(z) = L(g^*(z),\eta^*(z), z)\,,
\end{equation*}
for saddle points $(g^*(z), \eta^*(z))$. Next, we apply Theorem 4 in \cite{MilgromSegal2002} to the Lagrangian $L(g,\eta, z)$. For this note that for fixed $(g, \eta)$, $L(g,\eta, \cdot)$ is absolutely continuous, since the scoring function is continuously differentiable in $z$. Moreover, for each $z$, the set of saddle points is non-empty by \Cref{thm:KLUncertainty}. Also, there exists a non-negative and integrable function $b \colon \A \to [0, \infty)$ such that 
\begin{equation*}
    \big| \tfrac{\partial}{\partial z} L(g,\eta, z)\big|
    =
    \big|  \int \tfrac{\partial}{\partial z} S(z,y) g(y) \nu(dy)\big|
    \le b(z)\,,
\end{equation*}
where the inequality follows since the scoring function is locally Lipschitz continuous in $z$ (since it is continuously differentiable), and  $g$ is a density. Integrability of $b$ follows by the integrability assumption on the derivative of the scoring function.
Also, $ \tfrac{\partial}{\partial z} L(g,\eta, z) =  \tfrac{\partial}{\partial z}\mathcal{J}(g, z)$ is continuous in $g$ and $ L(g,\eta, z)$ equi-differentiable in $g$ and $\eta$. 
Thus the assumptions of Theorem 4 in \cite{MilgromSegal2002} are satisfied and it holds that 
\begin{equation}\label{eq:thm4-milgrom}
    J(z) = J(0) + \int_0^z  \tfrac{\partial}{\partial s'}L(g,\eta, s') \Big|_{g = g^*(s), \eta = \eta^*(s), s' = s}ds\,.
\end{equation}
Therefore, taking derivative with respect to $z$ of \eqref{eq:thm4-milgrom}, we have
\begin{align*}
    \tfrac{d}{dz}J(z) 
    &=
     \tfrac{\partial}{\partial z}L(g,\eta, z) \big|_{g = g^*(z), \eta = \eta^*(z)}
    \\
    &=
    \int \tfrac{\partial}{\partial z} S(z,y)g(y) \nu(dy)\big|_{g = g^*(z), \eta = \eta^*(z)}
    \\
     &=\frac{\E\left[\left\{\frac{\partial}{\partial z} S(z,Y) \right\}e^{\eta^*(z)S(z,Y)}\right]}{\E\left[e^{\eta^*(z)S(z,Y)}\right]} \,,
\end{align*}
where in the last equality we used that $g^*(z)$ and $\eta^*(z)$ are given in \Cref{thm:KLUncertainty}.

\underline{Part 2:}
To show that $\frac{d}{d z} J(z)$ crosses zero, we proceed as follows.
For fixed $\eta >0$, define $H^\eta(z,y) := e^{\eta S(z,y)}$  and $\Bar{H}^\eta(z) := \E[H^\eta(z,Y)]$. By \Cref{lemma-indentification}, $H^\eta(\cdot, \cdot)$ is strictly convex in $z$, locally Lipschitz continuous, and $[\;z_\eta^\mathrm{l}, z_\eta^\mathrm{u}\;]:= \argmin_{z\in \A} \Bar{H}^\eta(z)$ exists. Furthermore, by \Cref{lemma-indentification}, there exists an oriented identification function for $z_\eta^*$ that satisfies
\begin{equation*}
    k(z)\, W^\eta(z,y) =\eta  e^{\eta\, S(z,y)} \, \frac{\partial}{\partial z} S(z,y)\,,
\end{equation*}
for some $k(z) > 0 $, and for all $z \in \A$.
Since for each $\eta>0$, $W^\eta$ is an oriented identification function and as $k(\cdot) \ge 0$, we have that for all $z > z_\eta^\mathrm{u}$ 
\begin{equation*}
    k(z)\,  \E[W(z,Y)] = \E[e^{\eta S(z,Y)} \frac{\partial}{\partial z} S(z,Y)] > 0
\end{equation*}
and similarly for all $z < z_\eta^\mathrm{l}$ 
\begin{equation*}
    k(z)\, \E[W(z,Y)] = \E[e^{\eta S(z,Y)} \frac{\partial}{\partial z} S(z,Y)] < 0\,.
\end{equation*}
Since this holds for all $\eta>0$, the above equations also hold for $\eta^*$. Therefore
\begin{align*}
    &\tfrac{d}{dz} J(z) >0 \,, \quad \text{for all}\quad  z > z_{\eta^*}^\mathrm{u}\,, 
    \\[1em]
    &\tfrac{d}{dz} J(z) <0 \,, \quad \text{for all}  \quad z < z_{\eta^*}^\mathrm{l}\,,
    \quad \text{ and}
    \\[1em]
    &\tfrac{d}{dz} J(z)\big|_{z \in [z_{\eta^*}^\mathrm{l}, z_{\eta^*}^\mathrm{u}]} =0\, . 
\end{align*}
Therefore, $J(z)$ admits a (potentially interval valued) minima.

To show $ii)$, assume that $\argmin_{z \in \A} \E[ e^{v\, S(z,Y)}]$ is a singleton for all $v >0$. That is $z_{\eta^*}^l = z_{\eta^*}^u$ and is a singleton for all $\eta>0$. Which implies that $J(z)$ admits a unique minima.
\hfill\Halmos

We are able to alternatively write the REF in terms of the cgf of the scoring function. For this, we denote the derivative of the cgf of $Y\in\L$ by $K'_Y(t) := \frac{\partial}{\partial t} K_Y(t)$.

\begin{proposition}[Alternative representation]
The REF can be represented as:
\begin{equation*}
    T^{KL}(Y) = 
    \argmin_{z \in \A}\; K^\prime_{S(z,Y)}\big(\eta^*(z)\big)
    \,,
\end{equation*}
where for each $z \in \A$, $\eta^*(z)$ is the unique solution to 
\begin{equation}\label{eq:eta-cgf}
    \eta(z)  K'_{S(z,Y)}\big(\eta(z)\big) - K_{S(z,Y)}\big(\eta(z)\big) = \ep.
\end{equation}
\end{proposition}

\textit{Proof}
    Note that by definition, the cgf of $S(z,Y)$ is $K_{S(z,Y)(\eta)} = \log\left(\E[e^{\eta S(z,Y)}]\right)$. Then, we have 
    \begin{equation*}
        K'_{S(z,Y)}(\eta^*(z)) = \frac{\E[S(z,Y)e^{\eta(z) S(z,Y)}]}{\E[e^{\eta(z) S(z,Y)}]}.
\end{equation*}

    Moreover, \Cref{eq:eta-cgf} for $\eta(z)$ is equivalent to \Cref{eq:d}.  Thus, the two optimisation problems are equivalent.
\hfill \Halmos

\section{Choice of scoring function}
\label{sec: b-homog}

As elicitable functionals are defined as the $\argmin$ of an expected scoring function, the scoring function is not unique, indeed, there are infinitely many scoring functions. In this section, we discuss the family of homogeneous scoring function that naturally lead to illustration of the REF via Murphy diagrams.

\subsection{Families of $b$-homogeneous scoring functions}
To investigate the effect of the scoring function on the REF, we propose the use of $b$-homogeneous scoring functions as argued in \cite{Efron1991}, \cite{Patton2011}, and studied in \cite{NoldeZiegel2017}. 

\begin{definition}[$b$-Homogeneous Scores]
    A scoring function $S: \A' \times \R \longrightarrow [0,\infty)$, $\A' \subseteq \R$, is positively homogeneous of degree $b \in \R$, if 
    \begin{equation*}
        S(cz, cy) = c^b S(z, y)
    \end{equation*} 
    for all $(z,y) \in \A' \times \R$ and for all $c > 0$. 

   We say that a scoring function is positively homogeneous if there exists a $b \in \R$ such that it is positively homogeneous of degree $b$.
\end{definition}

These families of parameterised scoring functions retain useful properties of the elicitable functional, discussed next.

\begin{proposition}
\label{prop: properties}
    Let $S$ be a strictly consistent scoring function for $R$, then the following holds:

    \begin{enumerate}[label = $\roman*)$]
        \item If $S$ is a positive homogeneous scoring function, then $R$ and $\rfunc$ are positive homogeneous of degree $1$for $\A' \in\{ [0, \infty), (-\infty, 0], \R\}$.
        \item If $\A' = \R$ and $S(z-c, y) = S(z, y+c)$ for $c \in \R$, then $R$ and $\rfunc$ are translation invariant.
        \item If $S(y,y) = 0$ and $y \in \A'$, then $R(m) = m$ and $\rfunc(m) = m$ for all $m \in \R$.  In particular, $R(0) = 0$ and $\rfunc(0) = 0$.
    \end{enumerate}
\end{proposition}
\textit{Proof}
    \begin{enumerate}[label = $\roman*)$]
        \item The case for $R$  follows from \cite{FisslerPesenti2022}. Let $\A'\in\{ [0, \infty), (-\infty, 0], \R\}$. To show that it also holds for $\rfunc$, let $S$ be a positive homogeneous scoring function for $R$ of degree $b$.  Then, using the change of variable $z:= cw$, we obtain
    \begin{align*}
        \rfunc(cY) &= \argmin_{z \in \A'} \sup_{\Q \in \Qs} \E^\Q \left[S(z,cY)\right]\\
        &=  c\argmin_{w \in \A'} \sup_{\Q \in \Qs} \E^\Q \left[S(cw,cY)\right]\\
        &= c\argmin_{w \in \A'} \sup_{\Q \in \Qs} c^b \E^\Q \left[S(w,Y)\right]\\
        &= c\argmin_{w \in \A'} \sup_{\Q \in \Qs} \E^\Q \left[S(w,Y)\right]\\
        &= c \rfunc(Y)\,.
    \end{align*}
    \item Suppose that $S(z-c, y) = S(z, y+c)$ for $c \in \R$.  Fix $c \in \R$, then
    \begin{align*}
        \rfunc(Y+c) &= \argmin_{z \in \R} \sup_{\Q \in \Qs} \E^\Q \left[S(z,Y+c)\right]\\
        &= \argmin_{z \in \R} \sup_{\Q \in \Qs} \E^\Q \left[S(z-c,Y)\right]\\
        &= \argmin_{z \in \R} \left\{ \sup_{\Q \in \Qs} \E^\Q \left[S(z,Y)\right]\right\} +c \,\\
        &= \rfunc(Y) + c. 
    \end{align*}
    The statement for $R$ follows by setting $\ep = 0$.
    
    \item Assume that $S(y,y) = 0$ for $y \in \A'$ and let $m \in \R$. Then
    \begin{align*}
        \rfunc(m) 
        =
        \argmin_{z \in \A'} \sup_{\Q \in \Qs} \E^\Q\left[ S(z,m)\right]
        =
        \argmin_{z \in \A'} S(z,m)
        =
        m\,.
    \end{align*}
    The result for $R$ follows for the special case of $\ep = 0$.    \end{enumerate}
\hfill \Halmos

In the risk management literature, significant emphasis is placed on what is considered desirable properties of a risk measure.  Among these properties, positive homogeneity and translation invariance enjoy significant interest due to their inclusion into the notion of a coherent risk measure \citep{Artzner1999MF}.  Here, we show that we can translate these properties on the elicitable functional (risk measure) into properties on the corresponding scoring function, and that there is a relationship between these properties on the scoring function to their elicitable functional.

In particular, the positive homogeneity property, \Cref{prop: properties} case $i)$, allows for the rescaling of the rvs.  This has the financial interpretation of allowing for currency and unit conversions.  This also has practical implications, in that a practitioner can rescale $Y$, then estimate REF, for example to improve numerical performance, and then rescale the estimator back to the original magnitude, all using \Cref{prop: properties} case $i)$. 
Translation invariance, or cash invariance, is typically motivated by the risk-free property of cash assets. Meaning that by adding a constant value the risk of the random portfolio should be reduced by the same amount. A property that may be preserved by the REF.

\subsection{Murphy diagrams for robust elicitable functionals}

In this section we illustrate the REF on different functionals such as the mean, VaR, and expectiles using $b$-homogeneous scoring functions.  The $b$-homogeneous class of scoring functions allows practitioners to rescale losses via the homogeneity property of the REF, to improve numerical stability of the functional.  

\begin{proposition}[$b$-homogeneous scoring functions -- mean \citep{NoldeZiegel2017}]
\label{prop: hom mean}
The class of strictly consistent and $b$-homogeneous scoring functions $S_b^\E: [0, \infty)^2 \to \R $ for the mean satisfying $S(y,y)=0$ are given by any positive multiple of a member of the Patton family
\begin{equation}
    \label{eq:Patton}
    S_b^\E(z,y) = 
    \begin{cases}
    \frac{y^b - z^b}{b(b-1)}- \frac{z^{b-1}}{b-1}(y-z), & b\in\R\setminus \{0,1\},\\[0.5em]
    \frac{y}{z} - \log\left(\frac{y}{z}\right) - 1, & b=0,\\[0.5em]
    y\log\left(\frac{y}{z}\right) - (y-z), & b=1,
    \end{cases}
\end{equation}
where we require that $z,y>0$. 
\end{proposition}

Note that the squared-loss, $S(z,y) = (z-y)^2$, is recovered when $b = 2$ in the $b$-homogeneous scoring function for the mean in \Cref{eq:Patton}.

Introduced in \cite{EhmETAL2016}, a Murphy diagram is a graph used to display the effect of a scoring function's homogeneity parameter against the value of the functional. Here, we use this idea to plot the function $b \mapsto \overline{R}^{S_b}$. In the next examples, we plot the REF Murphy diagrams for the mean, VaR, and expectiles.

For the numerical examples, and as we work on $\L$, we consider right truncated rvs.  Right truncated rvs arise in financial and insurance contexts via financial options, limits on (re)insurance contracts, and maximal losses, e.g. if an insured asset is written off as a total loss. We also refer to \cite{AlbrecherBeirlantTeugels2017} for use of truncated and censored distributions in reinsurance settings.
Let $X$ be a random variable with pdf $g$ and cdf $G$. Then the right truncated random variable $Y := X \mid X \leq \bar{x}$ with truncation point $\bar{x}\in\R$, has pdf
\begin{equation*}
    f_Y(y \mid X \leq \bar{x}) = \frac{g(y)}{G(\bar{x})}\Id_{\{x \le \bar{x}\}}\,.
\end{equation*}
Furthermore, it holds that $F^{-1}(\alpha) = G^{-1}(\alpha G(\bar{x}))$, and $\E[Y] = \frac{\int^{\bar{x}}_0 x g(x) dx}{F(\bar{x})}$, whenever $Y \ge 0$ $\P$-a.s..

In the examples below, we consider right truncated exponential losses.  In particular, we truncate each exponential to its 95\% quantile i.e. we set the truncation point to  $\bar{x} := F^{-1}(0.95)$, where $F^{-1}$ is the quantile function of the exponential distribution. This corresponds to retaining 95\% of the support of the exponential distribution. We denote the above described distribution by TExp$(\lambda)$, where $\lambda$ is the parameter of the original exponential distribution.

\vspace{1em}
\begin{example}[Murphy diagrams for the mean]
\label{ex:mean}
We consider the mean functional and its REF, the REF mean, with the $b$-homogeneous scoring functions given in \Cref{prop: hom mean} for the Beta distribution. Here we use the convention that Beta($\beta_1$, $\beta_2$) has density $f(x) = \frac{\Gamma(\beta_1 + \beta_2)}{\Gamma(\beta_1)\Gamma(\beta_2)} x^{\beta_1-1} (1-x)^{\beta_2-1}$, where $\Gamma(\cdot)$ is the Gamma function. \Cref{fig:mean_beta_truncexp_b} displays the REF mean with varying uncertainty tolerances $\ep$ between $0$ and $0.5$, for the $\text{Beta}(\beta_1 =2, \beta_2 =2)$ baseline distribution  in the left panel of the figure and the TExp($2$) baseline distribution in the right panel. We observe that for the Beta distribution, for each $\ep$, the REF mean is increasing in the homogeneity degree $b$ and converges to the value 0.5, that is the mean of the Beta distribution. Furthermore, for fixed $b$, the larger the $\ep$, the smaller the REF mean. In particular, the REF mean is always less than the baseline mean. This is not observed in the right panel, which displays the REF mean for the TExp($2$) distribution, where the REF mean is always greater than the baseline mean of $0.4$. Moreover, for each $\ep$ the REF mean is increasing in the homogeneity degree $b$, and for fixed $b$ the REF mean is ordered in $\ep$. Therefore, we observe that depending on the underlying distribution the REFs can be greater or smaller than the baseline mean. 
\begin{figure}[t]
    \centering
    \includegraphics[width=0.7\textwidth]{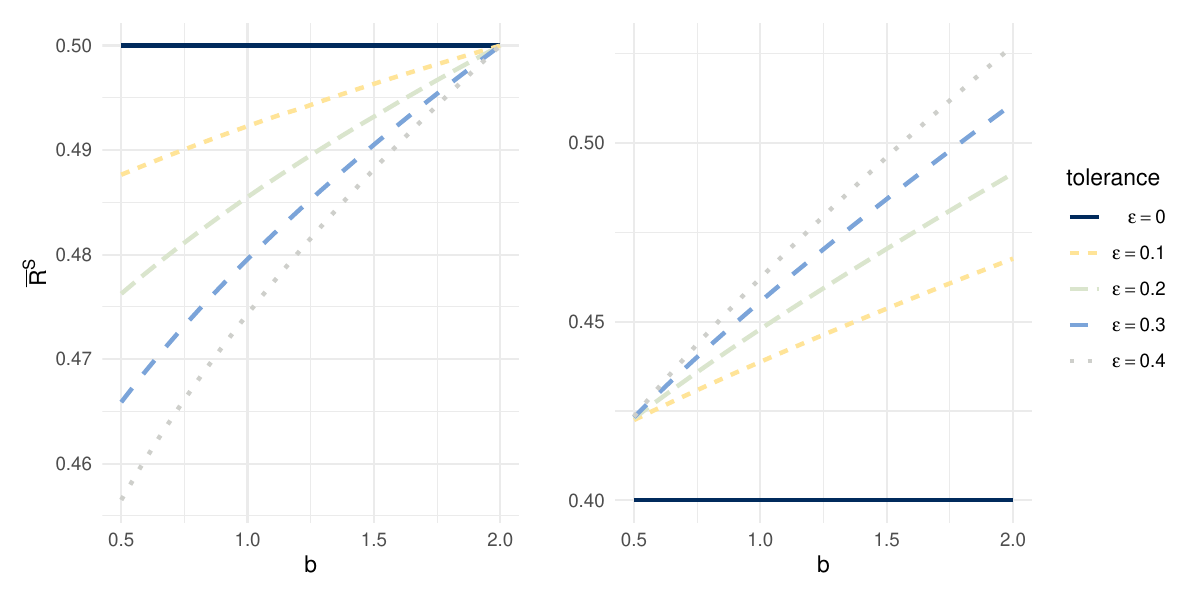}
    \caption{$\rfunc$ for varying $b$ parameter of the $b$-homogeneous mean scoring function for Beta(2,2) distribution (left), and TExp(2) (right).}
    \label{fig:mean_beta_truncexp_b}
\end{figure}
\Cref{fig:mean_beta_truncexp_b} also indicates that there is a trade-off between $\ep$ and $b$, indicating that the choice of tolerance $\ep$ should be influenced by the homogeneity degree $b$ of the scoring function.
\begin{figure}[h]
    \centering
    \includegraphics[width=0.7\textwidth]{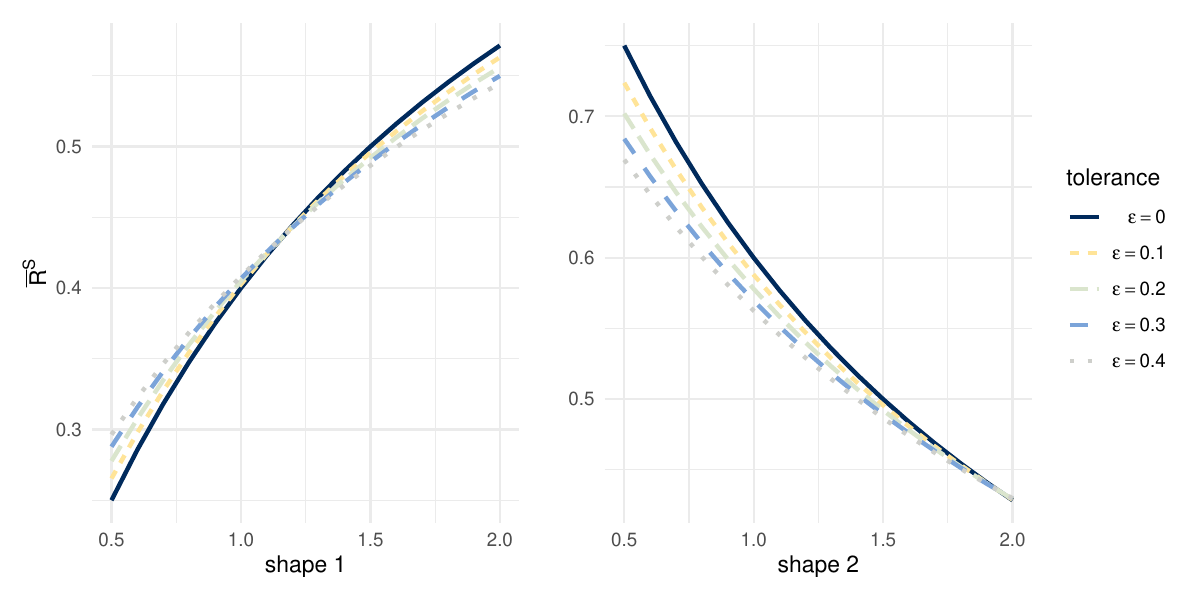}
    \caption{$\rfunc$ for varying shape parameters of the Beta distribution with $b = 1.5$ homogeneous mean scoring function, $\beta_2 \equiv 1.5$ (left), and $\beta_2 \equiv 1.5$ (right).}
    \label{fig:mean_beta_shape1_shape2}
\end{figure}
Next, we consider how the REF mean changes according to the parameters of the Beta distribution. In particular \Cref{fig:mean_beta_shape1_shape2} show the REF mean for fixed $b = 1.5$ against different values of the shape  parameters of the Beta distribution. The left panel displays variation with $\beta_1$ and the right panel with $b$, where the different lines correspond to different $\ep$. Interestingly, the REF  mean for different $\ep$ cross when plotted against the first parameter of the Beta distribution, whereas the REF  mean are ordered in $\ep$ for the second shape parameter $\beta_2$.
\end{example}
\vspace{1em}

Another elicitable functional of interest is the Value-at-Risk (VaR), also known as the quantile. 
\begin{definition}[Value-at-Risk]
    The \textbf{Value-at-Risk} at tolerance level $\alpha \in (0,1)$ is defined for $X \in \L$, as 
    \[
    \VaR_\alpha(X) = \inf\{x \in \R : \P(X\leq x)\geq \alpha\}\,.
    \]
\end{definition}

The VaR is well-known to be elicitable and its family of $b$-homogeneous scoring functions is recalled next.

\begin{proposition}[$b$-homogeneous scoring functions -- VaR \citep{NoldeZiegel2017}]
\label{prop: hom-score-var}
The class of strictly consistent and $b$-homogeneous scoring functions $S_b^\VaR: \R^2 \to [0,\infty)$ for $\VaR_\alpha$ satisfying $S(y,y)=0$ are given by 
\begin{equation*} 
    S_b^\VaR(z,y) = \left(\Id_{\{y \le z\}} - \alpha\right) \big(g(z) - g(y)  \big) \,,
\end{equation*}
where 
\begin{equation*}
g(y) = 
\begin{cases}
d_1y^b\Id_{\{y>0\}} - d_2|y|^b\Id_{\{y<0\}}  \qquad\qquad &\text{if} \quad b >0 \text{ and } y \in \R,
\\[0.5em]
d\log(y)  &\text{if} \quad b = 0 \text{ and } y >0,
\\[0.5em]
-dy^b   &\text{if} \quad b <0 \text{ and } y >0\,,
\end{cases}
\end{equation*}
for positive constants $d,d_1,d_2>0$.
\end{proposition}

The $b$-homogeneous scoring function for the VaR coincides with the pinball loss scoring function when $b = 1$.  For simplicity, we choose $d = d_1 = d_2 = 1$ in the following experiments.

\vspace{1em}
\begin{example}[Murphy diagrams for the VaR]
\label{ex:var}
Similarly to \Cref{ex:mean}, we plot the Murphy diagrams for the $\VaR$ functional for the TExp(2) baseline distribution. 
\begin{figure}[h]
    \centering
    \includegraphics[width=\textwidth]{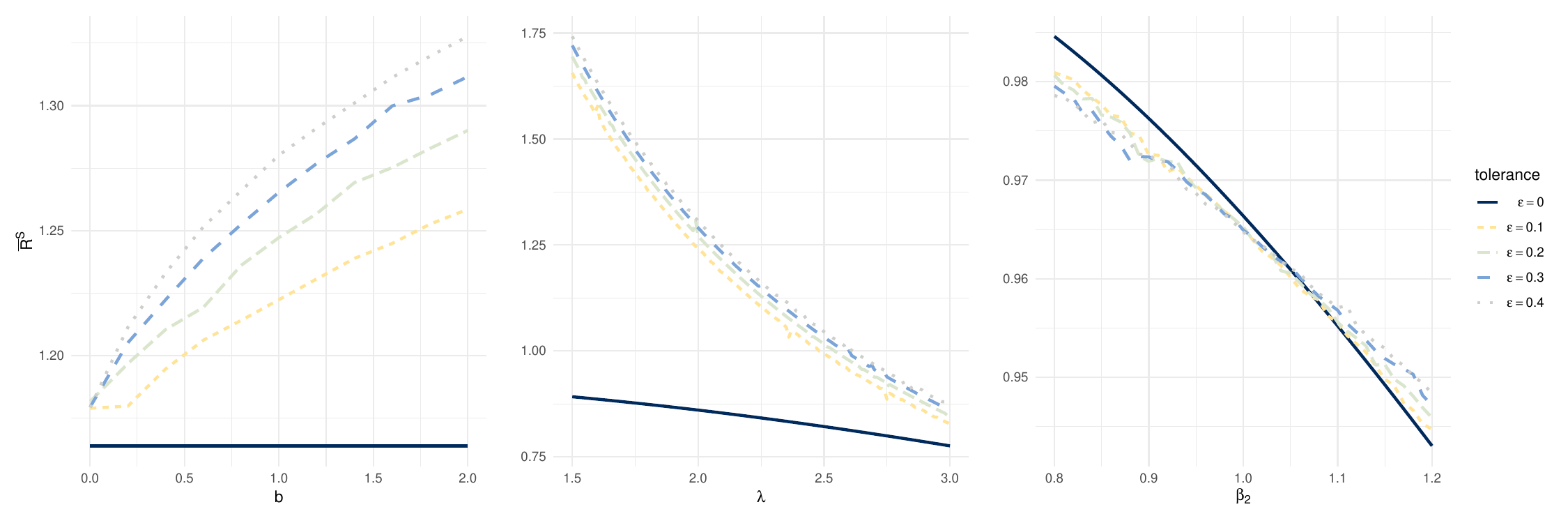}
    \caption{
    REF VaR at level $\alpha = 0.95$. Left: $\rfunc$ against the homogeneity parameter $b$ for baseline distribution TExp$(2)$. 
    Centre: $\rfunc$ against the parameter $\lambda$ of the baseline distribution TExp$(\lambda)$ and fixed $b = 1.5$. 
    Right:  $\rfunc$ against $\beta_2$ of the baseline distribution Beta($1.5, \beta_2$) and fixed $b = 1.5$.}
    
    \label{fig:var_truncexp_b_lambda}
\end{figure}
\Cref{fig:var_truncexp_b_lambda} displays in the left panel the REF $\VaR$s against the homogeneity degree $b$ for the baseline distribution TExp($\lambda$ = 2), on the centre panel against the parameter of the TExp($\lambda$) for fixed $b = 1.5$. The right most panel considers the Beta distribution as the baseline and plots the REF $\VaR$ against the second shape parameter, $\beta_2$ of the Beta distribution. We observe in all three panels that the REF $\VaR$s are ordered in $\ep$ with larger values for larger uncertainty tolerances $\ep$. Note that in the centre panel, for small $\lambda$, the difference between the REF $\VaR$ and the baseline VaR of the TExp($\lambda$) is significantly larger than the corresponding difference for large $\lambda$s. In the right panel, we observe that the REF $\VaR$ can be larger or smaller than the baseline $\VaR$.
\end{example}
\vspace{1em}

Another commonly considered functional is the expectile, which has been proposed for use in risk management in \cite{BelliniEtAl2014}.  Its elicitability is established in \cite{Gneiting2011}.

\begin{definition}[$\tau$-expectile]
    Let $\tau \in (0, 1)$, then the $\tau$-expectile, $e_\tau(\cdot)$, is the unique solution to
    \begin{equation*}
        \tau \E[(X- e_\tau(X))^+] = (1-\tau) \E[(X- e_\tau(X))^-]\;\,.
    \end{equation*}
\end{definition}

For $\tau \geq \frac{1}{2}$, the expectile is coherent in the sense of \cite{Artzner1999MF}. The family of $b$-homogeneous scoring functions for the expectile is as follows.

\begin{proposition}[$b$-homogeneous scoring functions -- expectile \citep{NoldeZiegel2017}]
\label{prop: expectile}

For $\tau \in (0,1)$, the strictly consistent $b$-homogeneous scoring function $S_b^e: [0, \infty)^2 \to \R$ corresponding to the $\tau$-expectile is given as
\begin{equation*}
    S_b^e(z,y)= \abs{1-\tau} \cdot S_b^\E(z,y),
\end{equation*}
where $S_b^\E(z,y)$ is the $b$-homogeneous scoring function for the mean as defined in \Cref{prop: hom mean}, and where $z,y > 0$.
\end{proposition}

\vspace{1em}
\begin{example}[Murphy diagrams for the expectile]
    \label{ex:expectile}
    We continue to consider the underlying distribution of the truncated exponential as described in \Cref{ex:var}. We calculate the $0.7$-expectile of the baseline TExp($3$) and TExp$(\lambda)$ numerically using a simulated sample of size 30,000.

\begin{figure}[h]
    \centering
    \includegraphics[width=0.7\textwidth]{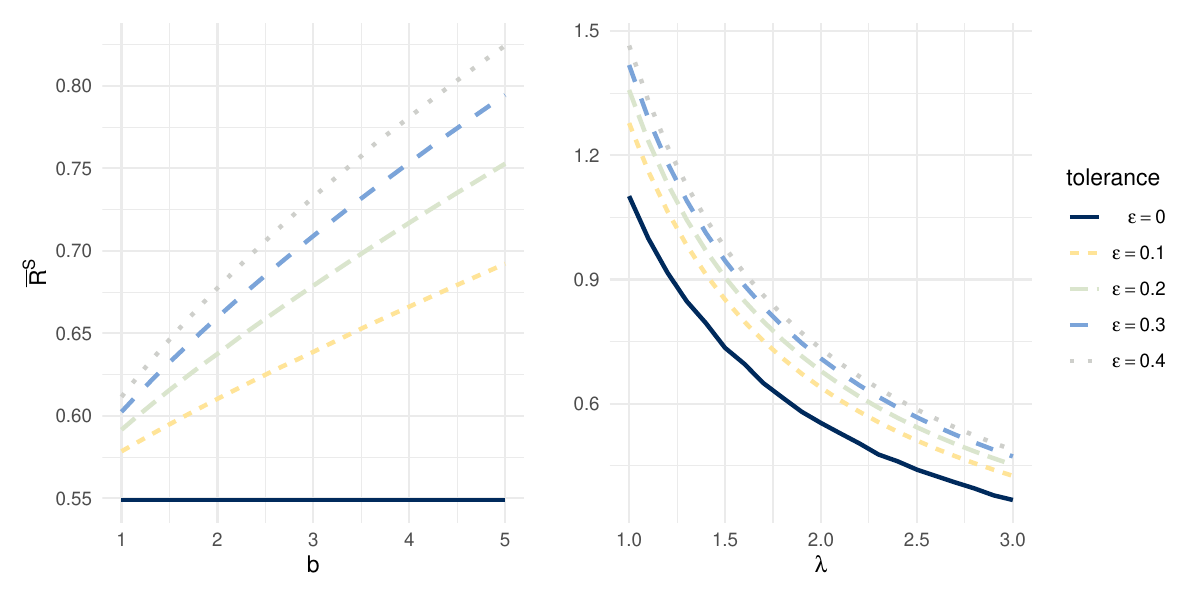}
    \caption{$\rfunc$ for varying $b$ parameter of the $b$-homogeneous $0.7$-expectile scoring function (left), and varying $\lambda$ with $b = 3$ homogeneous $0.7$-expectile scoring function (right), TExp$(2)$ distribution.}
    \label{fig:expectile_truncexp_b_lambda}
\end{figure}

The left panel of \Cref{fig:expectile_truncexp_b_lambda} displays the Murphy diagram of the $b$-homogeneous score for the expectile as defined in \Cref{prop: expectile}, and the right panel displays the robust expectile against the TExp parameter $\lambda$. The left panel in \Cref{fig:expectile_truncexp_b_lambda} exhibits similar behaviour as in previous examples for the mean and VaR. In particular, we have that for fixed $b$, the REF is increasing in tolerance $\ep$. This is again seen in the right panel, where for any fixed $\lambda$, the REF is increasing in $\ep$. 

\end{example}
\vspace{1em}

\section{Multivariate robust elicitability functional}
\label{sec: k-elicitable}

Many statistical functionals are not elicitable on their own, such as the variance, Range-Value-at-Risk, Expected Shortfall (ES), and distortion risk measures, \citep{Gneiting2011, KouPeng2016, WangZiegel2014, FisslerZiegel2019_RVaR}. However, they may be elicitable as multi-dimensional functionals, as is in the case of the pairs (mean, variance) and (VaR, ES) \citep{FisslerZiegelGneiting2016}. In this section, we generalise the REF to these instances where the functionals are elicitable in the multi-dimensional case. 

\subsection{The notion of $k$-dimensional elicitability}
In \cite{FisslerZiegel2016}, the authors establish the framework for $k$-consistency and $k$-elicitability, recalled next. 

\begin{definition}[$k$-elicitability \citep{FisslerZiegel2016}]

Let $\mathbf{\A} \subseteq \R^k$, $k \in \N$, be an action domain and $\mathbf{R}\colon \M \to \mathbf{\A}$ be a $k$-dimensional law-invariant functional.

\begin{enumerate}[label = $\roman*)$]
    \item A scoring function $S: \mathbf{\A} \times \R \to \R$ is \textbf{consistent} for a functional $\mathbf{R}$, if 
    \begin{equation}\label{eq:multi-scor-consistent}
     \int S(\mathbf{t},y) \,dF(y) \leq \int S(\mathbf{z}, y) \,dF(y),   
    \end{equation}
    for all $F \in \M$, $\mathbf{t} \in \mathbf{R}(F)$, and all $\mathbf{z} \in \mathbf{\A}$.

    \item 
    A scoring function $S$ is \textbf{strictly consistent} for a functional $\mathbf{R}$, if it is consistent and Equation \eqref{eq:multi-scor-consistent} holds with equality only if $\mathbf{z} \in \mathbf{R}(F)$.

    \item A functional $\mathbf{R}$ is \textbf{$k$-elicitable}, if there exists a strictly consistent scoring function $S$ for $\mathbf{R}$.
\end{enumerate}
\end{definition}

Similar to the one-dimensional case, elicitable functionals and strictly consistent scoring functions have a correspondence relationship. Indeed, a $k$-elicitable functional $\mathbf{R}(\cdot) := (R_1(\cdot), \ldots, R_k (\cdot))$ admits the representation
\begin{equation*}
   \big( R_1(Y), \ldots, R_k(Y)\big) = \argmin_{\bz \in \mathbf{\A}} \int S(\bz, y) \,dF_Y(y),
\end{equation*}
for all $F_Y \in \M$ and where $S$ is any strictly consistent scoring function for $\mathbf{R}$. 

Similar to univariate functionals, in the multi-dimensional setting, it is of interest to consider uncertainty in the baseline distribution or measure $\P$, and  we define the multi-dimensional REF as follows. Let $\mathbf{R}$ be an elicitable functional with strictly consistent scoring function $S$ and let $\ep \ge 0$. Then we define the \textbf{$k$-dimensional robust elicitable functional} (REF) evaluated at $Y \in \L$, by
\begin{equation}\label{opt:multiKL}
    \brfunc(Y) := \argmin_{\mathbf{z} \in \mathbf{\A}}\; \sup_{\Q \in \Qs_\ep}\; \E^\Q\left[S(\mathbf{z}, Y)\right]\,,
\end{equation}
where the uncertainty set $\Qs_\ep$ is given in \Cref{eq:uncertainty-set}.
As $\brfunc$ is $k$-dimensional, we write $\rfunc_i$ for the $i$-th component of $\brfunc$, thus $\brfunc(\cdot) := (\rfunc_1(\cdot), \ldots, \rfunc_k (\cdot))$.

The results of the univariate case \Cref{thm:KLUncertainty} follow readily into the multi-dimensional setting. For this we first generalise the assumption on the tolerance distance $\ep$.

\begin{assumption}[Maximal Kullback-Leibler distance]
\label{asm-k}
    Denote $p(\bz) := \P\big(S(\bz,Y) = \esssup S(\bz,Y)\big)$. Then one of the following holds:
\begin{enumerate}[label = $\roman*)$]
    \item    If $p(\bz) = 0$ for all $\z \in \mathbf{\A}$, then the tolerance distance satisfies $\ep \in[0, \infty)$. 

    \item If $p(\bz) > 0$ for some $\z \in \mathbf{\A}$, then the tolerance distance $\ep$ satisfies 
    $$0\le \ep <  \log \left(\tfrac{1}{p(\bz)}\right)\,
    \quad \text{for all} \quad \bz \in \mathbf{\A}\,.
    $$
\end{enumerate}
\end{assumption}

Under this condition, \Cref{thm:KLUncertainty} holds in the $k$-dimensional setting. 

\begin{corollary}
    [Kullback-Leibler Uncertainty]
\label{cor:multiKLUncertainty}

Let $\mathbf{R}$ be a $k$-elicitable functional, $S\colon \mathbf{\A}\times \R \to [0, \infty)$ be a strictly consistent scoring function for $\mathbf{R}$, and $\ep$ such that Assumption \ref{asm-k} is satisfied.
Then, the REF has representation
\begin{equation}\label{eq:kREF}
    \brfunc(Y) = 
    \argmin_{\mathbf{z} \in \mathbf{\A}}\; \frac{ \E\left[S(\mathbf{z}, Y) e^{\eta^*(\mathbf{z})S(\mathbf{z}, Y) }\right]}{\E\left[e^{\eta^*(\mathbf{z})S(\mathbf{z}, Y) }\right]}
    \,,
\end{equation}
where each $\mathbf{z}\in \mathbf{\A}$, $\eta^*(\mathbf{z}) \ge 0$ is the unique solution to $\ep = D_{KL}(\Q^{\eta(\mathbf{z})}~||~\P)$, with
\begin{equation*}
    \frac{d\Q^{\eta(\mathbf{z})}}{d\P}
    :=
    \frac{ e^{\eta(\mathbf{z})S(\mathbf{z}, Y) }}{\E\left[e^{\eta(\mathbf{z})S(\mathbf{z}, Y)  }\right]}\,.
\end{equation*}
\end{corollary}

\textit{Proof}
The proof is similar to that of \Cref{thm:KLUncertainty}. The inner optimisation problem of \eqref{eq:kREF} can be written as an optimisation problem over the density of $Y$ as follows
\begin{align*}
    \sup_{g\colon \R \to \R \; }\; \int S(\bz, y) g(y)dy\; \,,
    \quad \text{subject to}
    & \int \frac{g(y)}{f( y)}\log\left( \frac{g(y)}{f(y)}\right) f(y) \, dy\; \le \ep\,, 
    \\[0.5em]
    & \int g(y) dy\, = 1\,, 
    \quad \text{and}
    \\[0.5em]
    & g(y) \ge 0\,, \quad \text{for all } \; y \in \supp(Y)\,.
\end{align*}
This optimisation problem admits the Lagrangian, with Lagrange parameters $\eta_1, \eta_2 \geq 0$ and $\eta_3(y) \geq 0$ for all $y \in \supp(Y)$, 
\begin{equation*}
L(\eta_1, \eta_2, \eta_3, g) = \int \Big( -S(\bz, y) g(y) + \eta_1 g(y) \log\left(\frac{g(y)}{f(y)}\right) + \eta_2 g(y) - \eta_3(y) g(y)\Big) dy - \eta_1 \ep - \eta_2\,,
\end{equation*}
where $\eta_1$ is the Lagrange parameter for the KL constraint, $\eta_2$ is such that $g$ integrates to 1, and $\eta_3(y)$ are such that $g(y) \ge 0$, whenever $f(y)\ge0$, and $g(y) = 0$ otherwise.
Using similar steps as in \Cref{thm:KLUncertainty}, that is derive the Euler-Lagrange equation, solve it for $\frac{g(y)}{f(y)}$, and then impose $\eta_2, \eta_3$. Finally setting $\eta := \frac{1}{\eta_1}$ results in the change of measure 
\begin{equation*}
    \frac{d \Q^{\eta}}{d\P} := \frac{g(Y)}{f(Y)} = \frac{e^{\eta S(\bz, Y)}}{\E[e^{\eta S(\bz, Y)}]}\,.
\end{equation*}

Moreover, it holds that
\begin{equation*}
    D_{KL}(\Q^{\eta} ~||~ \P) = \eta K'_{S(\bz, Y)}(\eta) - \eta K_{S(\bz, Y)}(\eta)\,.
\end{equation*}

By similar arguments as in the proof of \Cref{thm:KLUncertainty}, the KL constraint is binding and $\eta^*$ is the unique solution to $\ep = D_{KL}(\Q^\eta ~||~ \P)$. As $\eta$ and $\Q^\eta$ both depend on $\bz \in \mathbf{\A}$, we make this explicit by writing $\eta(\bz)$ and $\Q^{\eta(\bz)}$ in the statement. 
\hfill \Halmos

\Cref{prop: properties} on the properties of the REF also hold in the $k$-dimensional setting.

\begin{proposition}
\label{prop: kproperties}
    Let $S: \mathbf{\A}' \times \R \to \R$, $\mathbf{\A}'\subseteq \R^k$, be a strictly consistent scoring function for a $k$-dimensional elicitable functional $\mathbf{R}$, then the following holds:

    \begin{enumerate}[label = $\roman*)$]
        \item If $S(c\, \bz, c\, y) = c^b S(\bz, y)$, where $c\, \bz: = (c z_1, \ldots, cz_k)$, for all $y \in \R$, $c > 0$, and  for some $b \geq 0$. Then $\mathbf{R}$ and $\brfunc$ are positive homogeneous of degree $1$ for $\mathbf{\A}' = (-\infty, 0]^k, [0, \infty)^k$ or $\R^k$.
        \item If $\mathbf{\A}' = \R^k$ and $S(\bz -\mathbf{c}, y) = S(\bz, y+c)$ for $c \in \R$, $\mathbf{c} := (c, \ldots, c)$ of length $k$, then $\mathbf{R}$ and $\brfunc$ are translation invariant.
        \item If $S(\mathbf{y},y) = 0$, where $\mathbf{y} := (y, \ldots, y)\in \mathbf{\A}'$, then $\mathbf{R}(m) = m$ and $\brfunc(m) = m$ for all $m \in \R$.  In particular, $\mathbf{R}(0) = 0$ and $\brfunc(0) = 0$.
    \end{enumerate}
\end{proposition}

\textit{Proof} The proof follows using similar arguments as in the proof of \Cref{prop: properties}. \hfill \Halmos

A functional of major interest in the context of risk management is the ES. 

\begin{definition}[Expected Shortfall]
    The \textbf{Expected Shortfall}, also known as the conditional value-at-risk (CVaR), at tolerance level $\alpha\in [0,1)$ for $X \in \L$ is defined as 
    \[
    \ES_\alpha(X) = \frac{1}{1-\alpha}\int_\alpha^1 \VaR_q(X) dq\;.
    \]
\end{definition}

Though it is known to not be elicitable alone \citep{Gneiting2011}, the pair ($\VaR_\alpha$, $\ES_\alpha$) is $2$-elicitable, as shown in \cite{FisslerZiegelGneiting2016}.  While many scoring functions exist for this pair, we refer to Theorem C.3 in supplemental of \cite{NoldeZiegel2017} for the existence of a $b$-homogeneous scoring function for $(\VaR_\alpha, \ES_\alpha)$.

\begin{proposition}[$b$-homogeneous scoring function for (VaR, ES) \citep{NoldeZiegel2017}]

    For $\alpha \in (0,1)$, the $2$-elicitable functional $(\VaR_\alpha, \ES_\alpha)$ has corresponding scoring functions $S: \R^3 \to \R$ of the form
    \begin{equation}
    \label{eq:b_homog_VaR_ES}
        S(z_1, z_2, y) = \Id_{\{y > z_1\}} \big( -G_1(z_1) + G_1(y) - G_2(z_2)(z_1-y)\big) + (1 - \alpha) \,\big(G_1(z_1) (z_2 - z_1) + \mathcal{G}_2(z_2)\big)\,,
    \end{equation}
    
    where $G_1$ is increasing, and $\mathcal{G}_2$ is twice differentiable, strictly increasing, strictly concave and $\mathcal{G}_2' = G_2$. If

\begin{enumerate}[label = $\roman*)$]
    \item $b \in (0,1)$, the only positive homogeneous scoring functions of degree $b$ and of the form in \Cref{eq:b_homog_VaR_ES} are obtained by $G_1(x) = (d_1 \Id\{x \geq 0\} - d_2 \Id\{x < 0\}) \abs{x}^b - c_0$ and $\mathcal{G}_2(x) = c_1 x^b + c_0$, $x>0$ where $c_0 \in \R$, $d_1, d_2 \geq 0$, and $c_1 > 0$.
    
    \item $b \in (-\infty, 0)$, the only positive homogeneous scoring function of degree $b$ and of the form in \Cref{eq:b_homog_VaR_ES} are obtained by $G_1 (x) = -c_0$ and $\mathcal{G}_2 (x) = -c_1 x^b + c_0$, $x > 0$ where $c_0 \in \R$ and $c_1 > 0$.
    \item $b \in \{0\}\cup (1, +\infty)$, there are no positively homogeneous scoring functions of the form in \Cref{eq:b_homog_VaR_ES} of degree $b = 0$ or $b \geq 1$.
\end{enumerate}
\end{proposition}
We refer to the supplemental of \cite{NoldeZiegel2017} for further details and discussions.

In the next section, we discuss an application of jointly robustifying the (VaR, ES) to a reinsurance application. 

\subsection{Reinsurance application}\label{subsec-reinsurance}
We consider a reinsurance company who aims to assess the risk associated with its losses. In particular, we are interested in an reinsurance company who has underwritten losses stemming from different insurance companies. Specifically, the insurance-reinsurance market consists of three insurers and one reinsurer. Each insurance company $k \in \{1, 2,3\}$ purchases reinsurance on their business line $X_k$ with deductible $d_k$ and limit $l_k$. Thus, the total reinsurance loss is 
\begin{align*}
    Y &=
    \sum_{k  = 1}^3 \; \min\big\{\left(X_k - d_k\right)_+, \;  l_k\big\}\,.
\end{align*}
The reinsurer is covering losses between the 60\% and the 80\% quantile for insurer 1 and 2, and the losses between the 85\% and the 95\% quantiles of insurer 3, i.e. 
\begin{align*}
    d_k :&= F_{X_k}^{-1}(0.6)\phantom{5} \quad \text{and} \quad l_k := F_{X_k}^{-1}(0.8)\,,
    \quad k = 1,2\,, \quad \text{and} 
    \\
    d_3 :&= F_{X_3}^{-1}(0.85) \quad \text{and} \quad l_3 := F_{X_3}^{-1}(0.95)\,.    
\end{align*}
The insurance losses  $(X_1, X_2, X_3)$ have marginal distributions described in \Cref{tab: ex: reinsurance} and are dependent through a $t$-copula with 4 degrees of freedom and correlation matrix
\begin{equation*}
    R = 
    \left(
    \quad
    \begin{array}{@{} 
    S[table-format=1.3]
    S[table-format=1.3]
    S[table-format=1.1] @{}}
    1  & 0.2 &  0\\
    0.2 & 1 &  0.8 \\
    0 & 0.8 & 1  \\
    \end{array}
    \quad
    \right)\,.
\end{equation*}

\begin{table}[h]
\centering
\caption{Distributional assumptions of risk factors of the reinsurance example.}\label{tab: ex: reinsurance}
\begin{tabular}{ c @{\hspace{2.5em}}  c @{\hspace{2em}} c c}
Risk factor & Distribution  & Mean & Std \\
\toprule 
\toprule 
$X_1$ &   Log-Normal$\, \left(4.58, ~ 0.19^2\right)$ & \multicolumn{1}{d{3.2}}{100} &  \multicolumn{1}{d{3.2}}{20}\\[0.5em]
$X_2$ &   Log-Normal$\, \left(4.98, ~ 0.23^2\right)$ & \multicolumn{1}{d{3.2}}{150} &   \multicolumn{1}{d{3.2}}{35}\\[0.5em]
$X_3$ &   Pareto$\, \left(147.52, ~ 60.65\right)$ & \multicolumn{1}{d{3.2}}{150} &   \multicolumn{1}{d{3.2}}{40}\\[0.5em]
\bottomrule
\bottomrule
\end{tabular}
\end{table}

\begin{figure}[h]
    \centering
    \includegraphics[width=0.7\textwidth]{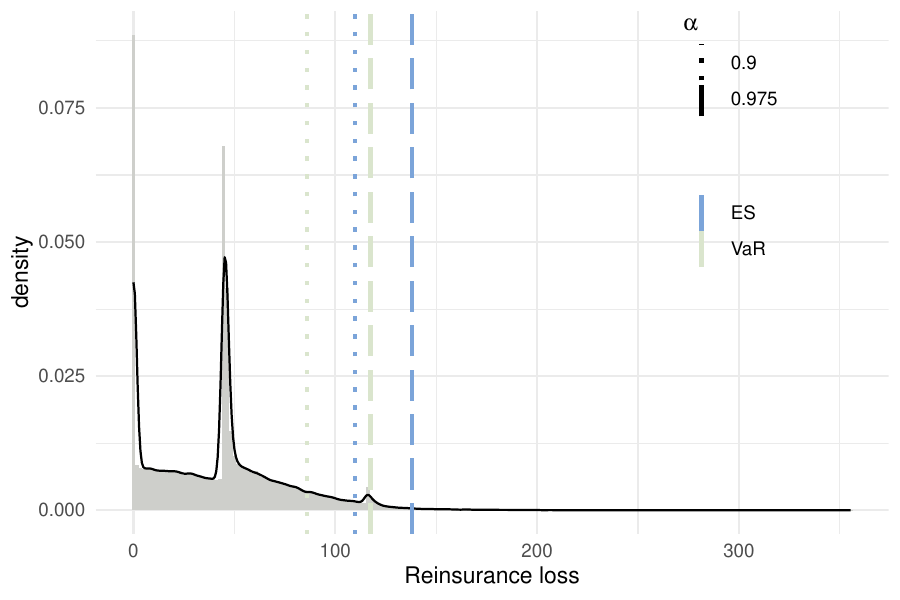}
    \caption{Smoothed density for $n = 100,000$ simulated reinsurance losses as described in \Cref{tab: ex: reinsurance}, with $\VaR_\alpha$ and $\ES_\alpha$ for $\alpha = 90\%, 97.5\%$. }
    \label{fig:reinsurance_density}
\end{figure}

\Cref{fig:reinsurance_density} displays the histogram of the total reinsurance losses $Y$, stemming from $n = 100,000$ samples, and the estimated baseline $\VaR_\alpha(Y)$ and $\ES_\alpha(Y)$ for $\alpha = 0.9,\; 0.975$.  A smoothed kernel density is displayed as an approximation of the density. In the simulated dataset, the reinsurer's losses are bounded by the sum of the insurers' limit, and the maximal loss for the reinsurer is $448.15$. The baseline/empirical $\VaR$s are  found to be 86.01 and 109.72 and the baseline/empirical $\ES$s are 117.40 and 138.13, for $\alpha = 90\%, \, 97.5\%$ respectively.

Next, we generate a sample of size $n = 10,000$ of reinsurance losses, and scale the losses by a factor of $0.01$.  We are able to perform this scaling by using a $b$-homogeneous scoring function for the pair $(\VaR_\alpha, \ES_\alpha)$ and the homogeneity property from \Cref{prop: kproperties}. 

\begin{figure}[h]
    \centering
    \includegraphics[width=0.7\textwidth]{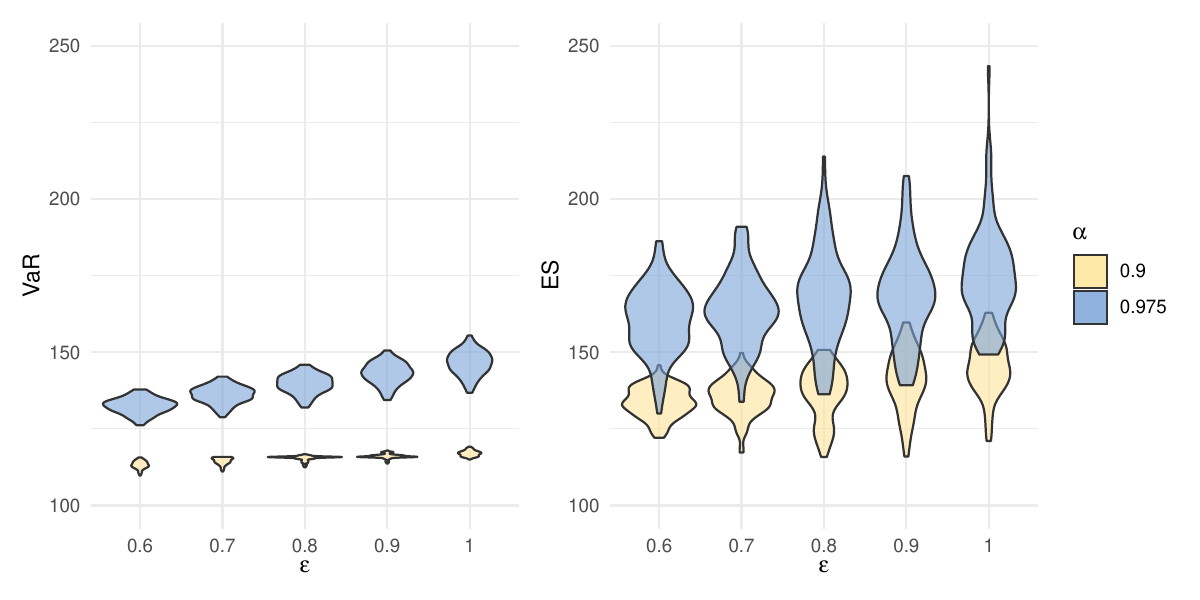}
    \caption{Densities of robust $\VaR$ (left) and robust $\ES$ (right) of simulated reinsurance losses for varying $\ep$ and $\alpha$. Reference values for VaR are 86.01 and 109.72 and for ES are 117.40 and 138.13, with $\alpha = 90\%, \, 97.5\%$ respectively.}
    \label{fig:reinsurance_var_es}
\end{figure}

For this sample, we calculate the joint REF $(\VaR, \ES)$. We do this $N = 100$ times to illustrate the density of the joint REF $\VaR$ and $\ES$.  To handle quantile-crossing, we reject loss samples that result in $\VaR_\alpha > \ES_\alpha$ for any $\ep = 0.6, 0.7, 0.8, 0.9$, $\alpha = 0.9, 0.975$. \Cref{fig:reinsurance_var_es} displays violin diagrams for the joint REF VaR and ES, for the different $\ep$'s and $\alpha$'s. 
We find that the joint REF VaR and ES are ordered in $\ep$, mirroring the behaviour of the one-dimensional REFs.  We also observe that the variance of the REF VaR is significantly smaller than that of the REF ES when found jointly.

\section{Application to robust regression}
\label{sec: rob-reg}

In this section, we extend the REF framework to a regression setting. Distributionally robust regression has been of increasing interest in the machine learning and statistics communities.
In the field of statistics, emphasis is placed on extreme value detection and methods that are not too sensitive to extreme values. Significant theory has been developed on this topic, see e.g.  \cite{RousseeuwLeroy2005} for a comprehensive treatment. These methodologies are now widely applicable to problems in machine learning, though their goals differ. For example, robustness techniques can be used to counter adversarial tests against machine learning algorithms.  Specifically for robust regression, the authors \cite{ShafieezadehEtAl2015}, consider robust logistic regression with uncertainty on both the covariates and the response quantified via a Wasserstein ball, and, in an attempt to mitigate adversarial attacks, \cite{ChenPaschalidis2018} minimise the worst-case absolute residuals of a discrete regression problem over a Wasserstein ball.  Data contamination in DRO problems have likewise been studied, see \cite{XuZhang2021}, where authors find sufficient conditions for quantitative robustness of a statistical estimator of DRO models.

The motivation for robustness in risk management differs from these communities.  Of concern are extreme events, where data may be sparse due to the rarity of events, which a risk measure should ideally be  capturing.

\subsection{Robust regression coefficients}
Here we propose an approach where we consider distributional uncertainty jointly in both the covariate and the response variable, and where the uncertainty is characterised via the KL-divergence. For this let $\bX:=(X_1, \ldots, X_m)$ be the $m$-dimensional covariates such that each component is in $\L$, i.e., $X_k \in \L$, for all $k = 1\ldots, m$, and let $\Y \in \L$ be a univariate response.

For an elicitable functional $R$ with strictly consistent scoring function $S$, we make the classical regression assumption that 
\begin{equation*}
    R(Y|\X = \x) = \beta_1 x_1 + \ldots +  \beta_m x_m\,,
\end{equation*}
where $R(Y|\X = x)$ denotes the functional $R$ evaluated on the conditional cdf of $Y$ given $\bX = \x$. The parameters $\bbeta:=(\beta_1, \ldots, \beta_m)$, the regression coefficients, are estimated via solving the sample-version of the following minimisation problem
\begin{equation}\label{eq:regression}
    \hat{\bbeta} = \argmin_{\bbeta \in \R^m} \E\big[S(\bbeta^{\intercal} \bX, Y)\big]\;.
\end{equation}

As we allow $\bbeta \in \R^m$, in this section, we only consider scoring functions defined on $\R^2$, i.e. $S: \R^2 \to [0, \infty)$. Moreover, for simplicity, we assume that the functional $R(Y|\bX= x)$ is linear in the covariates, though the results can be adapted to include link functions.
For the choice of the mean functional $R$ and  the squared loss $S(z,y) = (y-z)^2$, we recover the usual linear regression. The classical quantile regression follows by setting the scoring function to be the pinball loss, i.e., $S(z,y) = \big(\Id_{\{y \le z\}} - \alpha\big) (z-y)$, which is strictly consistent for the $\alpha$-quantile, $\alpha \in (0,1)$. Similarly one can obtain expectile and ES regression.

We propose to robustify the regression coefficients $\hat{\bbeta}$ of \Cref{eq:regression} by accounting for uncertainty in the joint distribution between the response and the covariates, that is of $(\bX, Y)$.

\begin{definition}[Robust regression coefficients]
    Let $R$ be an elicitable functional with strictly consistent scoring function $S\colon \R^2 \to [0, \infty)$ and $\ep\ge 0$. Then the \textbf{robust regression coefficients} are given by
\begin{equation}\label{eq:rob-coef}
    \breg := \argmin_{\bbeta \in \R^m}\; \sup_{\Q \in \Qs_\ep}\; \E^\Q\left[S(\bbeta^\intercal \mathbf{X}, Y)\right]\,,
\end{equation}
where the uncertainty set $\Qs_\ep$ is given in \Cref{eq:uncertainty-set}.
\end{definition}

The representation of the inner optimisation problem in \eqref{eq:rob-coef} also holds in the robust regression setting under a similar assumption on the tolerance distance $\ep$.

\begin{assumption}[Maximal Kullback-Leibler distance]
\label{asm-reg}

Define $p(\bbeta) := \P\big(S(\bbeta^\intercal \bX,Y) = \esssup S(\bbeta^\intercal \bX,Y)\big)$. Then one of the following holds:
\begin{enumerate}[label = $\roman*)$]
    \item    If $p(\bbeta) = 0$ for all $\bbeta \in \R^m$, then the tolerance distance satisfies $\ep \in[0, \infty)$. 

    \item If $p(\bbeta) > 0$ for some $\bbeta \in \R^m$, then the tolerance distance $\ep$ satisfies 
    $$0\le \ep <  \log \left(\tfrac{1}{p(\bbeta)}\right)\,
    \quad \text{for all} \quad \bbeta \in \R^m\,.
    $$
\end{enumerate}
\end{assumption}

\begin{corollary}
    Let $S$ be a strictly consistent scoring function for $R$, and $\ep$ such that Assumption \ref{asm-reg} is satisfied.  Then, for $m$-dimensional covariates $\bX$, satisfying $X_k \in \L$, $k = 1, \ldots, m$, and a univariate response $Y \in \L$, the robust regression coefficients have representation
    \begin{equation*}
        \breg = \argmin_{\bbeta \in \R^m}\frac{ \E\left[S(\bbeta^\intercal \bX, Y) e^{\eta^*(\bbeta)S(\bbeta^\intercal \bX, Y) }\right]}{\E\left[e^{\eta^*(\bbeta)S(\bbeta^\intercal \bX, Y) }\right]}\,,
    \end{equation*}
where for each $\bbeta\in\R^m$, $\eta^*(\bbeta) \ge 0$ is the unique solution to $\ep = D_{KL}(\Q^{\eta(\bbeta)}~||~\P)$ with 
\begin{equation*}
\frac{d\Q^{\eta(\bbeta)}}{d\P} := \frac{ e^{\eta(\bbeta)S(\bbeta^\intercal \bX, Y) }}{\E\left[e^{\eta(\bbeta)S(\bbeta^\intercal \bX, Y)  }\right]}\,.    
\end{equation*}
\end{corollary}

\textit{Proof}
The proof is similar to that of \Cref{thm:KLUncertainty}. For simplicity, we assume that $(\bX,Y)$ has joint pdf $f$ under the baseline measure $\P$. Then, the inner optimisation problem of \eqref{eq:rob-coef} can be written as an optimisation problem over the joint density of $(\X, Y)$ as follows
\begin{align*}
    \sup_{g\colon \R^{m+1}  \to \R \; }\; \int \int S(\bbeta^{\intercal} \mathbf{x}, y) g(\mathbf{x}, y)dy\; & d\mathbf{x}\,,
    \quad \text{subject to}\\
    & \int \int \frac{g(\mathbf{x}, y)}{f(\mathbf{x}, y)}\log\left( \frac{g(\mathbf{x}, y)}{f(\mathbf{x}, y)}\right) f(\mathbf{x}, y) \, dy\; d\mathbf{x}  \le \ep\,, 
    \\[0.5em]
    & \int \int g(\mathbf{x}, y) dy\, d\mathbf{x} = 1\,, 
    \quad \text{and}
    \\[0.5em]
    & g(\mathbf{x}, y) \ge 0\,, \quad \text{for all } \mathbf{x} \in \R^m, \; y \in \supp(Y)\,.
\end{align*}
This optimisation problem admits the Lagrangian, with Lagrange parameters $\eta_1, \eta_2 \geq 0$ and $\eta_3(\bx, y) \geq 0$ for all $\bx \in \R^m, y \in \supp(Y)$, 
\begin{align*}
     L(\eta_1, \eta_2, \eta_3, g) 
     &= 
     \int \int \Big( -S(\bbeta^\intercal \bx, y) g(\bx, y) + \eta_1 g(\bx, y) \log\left(\frac{g(\bx, y)}{f(\bx, y)}\right) 
     \\
     & \qquad\qquad + \eta_2 g(\bx, y) - \eta_3(\bx, y) g(\bx, y)\Big) dy\; d\bx\\
     & \qquad \qquad - \eta_1 \ep - \eta_2\,,
\end{align*}
where $\eta_1$ is the Lagrange parameter for the KL constraint, $\eta_2$ is such that $g$ integrates to 1, and $\eta_3(\x, y)$ are such that $g(\x, y)\ge0$, whenever $f(\x, y)\ge0$, and $g(\x, y) = 0$ otherwise.
Using similar steps as in \Cref{thm:KLUncertainty}, that is derive the Euler-Lagrange equation, solve it for $\frac{g(\x, y)}{f(\x, y)}$, and then impose $\eta_2, \eta_3$. Finally setting $\eta := \frac{1}{\eta_1}$ results in the change of measure 
\begin{equation*}
    \frac{d \Q^{\eta}}{d\P} := \frac{g(\bX, Y)}{f(\bX, Y)} = \frac{e^{\eta S(\bbeta^\intercal \bX, Y)}}{\E[e^{\eta S(\bbeta^\intercal \bX, Y)}]}\,.
\end{equation*}

Moreover, it holds that
\begin{equation*}
    D_{KL}(\Q^{\eta} ~||~ \P) = \eta K'_{S(\bbeta^\intercal \bX, Y)}(\eta) - \eta K_{S(\bbeta^\intercal \bX, Y)}(\eta)\,.
\end{equation*}

By the similar arguments as in the proof of \Cref{thm:KLUncertainty}, the KL constraint is binding and $\eta^*$ is the unique solution to $\ep = D_{KL}(\Q^\eta ~||~ \P)$. As $\eta$ and $\Q^\eta$ both depend on $\bbeta \in \R^m$, we make this explicit by writing $\eta(\bbeta)$ and $\Q^{\eta(\bbeta)}$ in the statement. 
\hfill \Halmos

We consider a numerical case study where we compare three different models for the joint dependence structure of a bivariate covariate and response pair $(X, Y)$, as detailed below. Here the dimension of the covariates is 1, i.e., $m = 1$.  Motivated by a dataset being contaminated with confounding or extreme values, we prescribe $(X, Y)$ as having a Gumbel($5$) copula with uniform marginals, as the ``reference'' samples without any confounding values (model A). The choice of Gumbel($5$) copula yields a Kendall's tau of $0.8$, giving a suitable linear relationship to test linear regression. The ``contaminated'' models (models B and C) have additional to the data points of model A, samples of $(\tilde{X}, \tilde{Y})$, that we identify as confounding values, where $(\tilde{X}, \tilde{Y})$ have uniform marginal distributions and an independent copula.

\vspace{1em}
\begin{example}[Data Contamination]
\label{ex:rlr}
In particular, the dataset we consider is constructed in the following way: we sample from the Gumbel($5$) copula to form the uncontaminated dataset in model A, then augment model A with 4 independent extreme values to result in model B, and then again augment model B with 4 additional extreme values to generate model C.  The model B consists of 9\% independent extreme values, and model C of 18\%. We fit a robust linear regression, with varying tolerances, to these 3 models, using the squared loss scoring function $S(z,y) = (z-y)^2$.  We further calculate the traditional linear regression, which coincides with tolerance $\ep = 0$.  
\begin{figure}[h]
    \centering
    \includegraphics[width=0.8\textwidth]{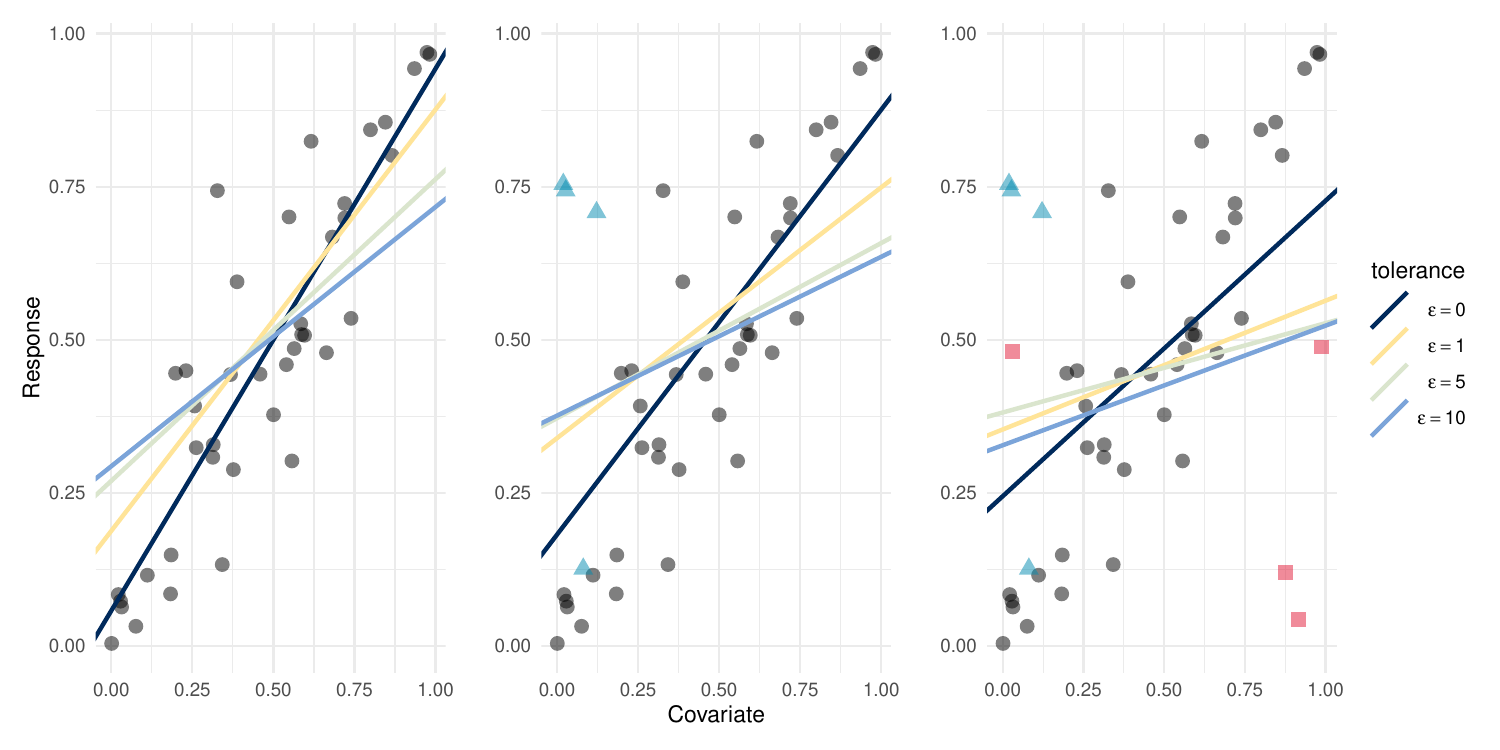}
    \caption{Fitted robust linear regressions lines for models A, B, C (left to right) and for different $\ep = 0, 1, 5, 10$. The blue pyramid and the red squares correspond to the extreme values in models B and C.}
    \label{fig:rlr}
\end{figure}
\Cref{fig:rlr} displays the datasets for models A, B, and C in the panels from left to right, and the robust regression lines for varying tolerances $\ep$. In the leftmost panel, we have the uncontaminated dataset model A, where we find that the robust linear regression has steeper regression lines for smaller tolerances. The centre panel displays the data and linear regression lines from the mildly contaminated model B.  The regression lines from this centre panel behave similarly to those of model A, in that steeper regression lines correspond to smaller $\ep$, though the confounding of the sample reduces the  overall steepness of the regression lines. In the left and middle panels, the slope of the robust regression decreases as the uncertainty tolerance $\ep$ increases, which is different from the right panel where we do not observe a clear ordering. The right panel displays the robust linear regression of the most contaminated dataset, model C.  In this scenario, the robust linear regression is significantly flatter for all values of $\ep$ compared to the linear regression lines with $\ep = 0$. Clearly, including extreme values reduces the slope of the regression line. Moreover, we observe that the larger the tolerance distance $\ep$, meaning as we allow for more uncertainty, the flatter the regression lines. 

For completeness, we report in \Cref{tab: ex: RLR} the regression coefficients and mean squared errors (MSEs) of the robust linear regressions of models A, B, and C. As observed in \Cref{fig:rlr}, the larger $\ep$ the smaller the slope of the regression line, i.e. $\beta_1$. 

\begin{table}[ht]
    \centering
    \caption{Results of robust linear regression for models A, B, and C. The parameter $\beta_0$ corresponds to the intercept, $\beta_1$ is the slope of the regression line, and MSE the mean squared error.\\}
    \label{tab: ex: RLR}
    \begin{tabular}{ c @{\hspace{2.5em}} c c c l @{\hspace{2.5em}} c c c l @{\hspace{2.5em}} c c c}
     &\multicolumn{3}{c}{Model A} &&  \multicolumn{3}{c}{Model B} && \multicolumn{3}{c}{Model C}
    \\
    \cmidrule(lr){2-4} \cmidrule(lr){6-8} \cmidrule(lr){10-12}
    \\
     $\ep$ & $\beta_0$  & $\beta_1 $& MSE & &   $\beta_0$  & $\beta_1 $& MSE & & $\beta_0$  & $\beta_1 $& MSE \\
    \toprule 
    \toprule
     0 & 0.05 & 0.88 & 0.016 & & 0.18 & 0.69 & 0.036 & & 0.24 & 0.48 & 0.054\\[0.5em]
     1 & 0.18 & 0.68 & 0.020 & & 0.40 & 0.33 & 0.044 & & 0.35 & 0.21 & 0.062\\[0.5em]
     5 & 0.26 & 0.49 & 0.028 & & 0.37 & 0.28 & 0.051 & & 0.38 & 0.14 & 0.065 \\[0.5em]
     10 & 0.29 & 0.42 & 0.030 & & 0.37 & 0.25 &  0.051 & & 0.32 & 0.19 &  0.079\\[0.5em]
    \bottomrule
    \bottomrule
    \end{tabular}
    \end{table}
\end{example}
\vspace{1em}

\begin{example}[Sensitivity to sample size]
Another situation in which one may want to robustify linear regression is when the data sample is sparse, e.g. to introduce uncertainty to prevent model overfit. Here, we consider the behaviour of the robust regression for increasing sample sizes. We generate covariate and response pairs having Gumbel($5$) copula and uniform marginals, i.e. Model A. The first dataset consists of $40$ samples, then we increase the dataset to $80$ samples by adding another $40$ samples, resulting in the second dataset. Finally, for the third dataset, we add another $40$ samples to obtain a sample size of $120$. 
\Cref{fig:rlrn} displays model A for sample sizes $n = 40, 80, 120$, from left to the right panel.
\begin{figure}[h]
    \centering
    \includegraphics[width=0.8\textwidth]{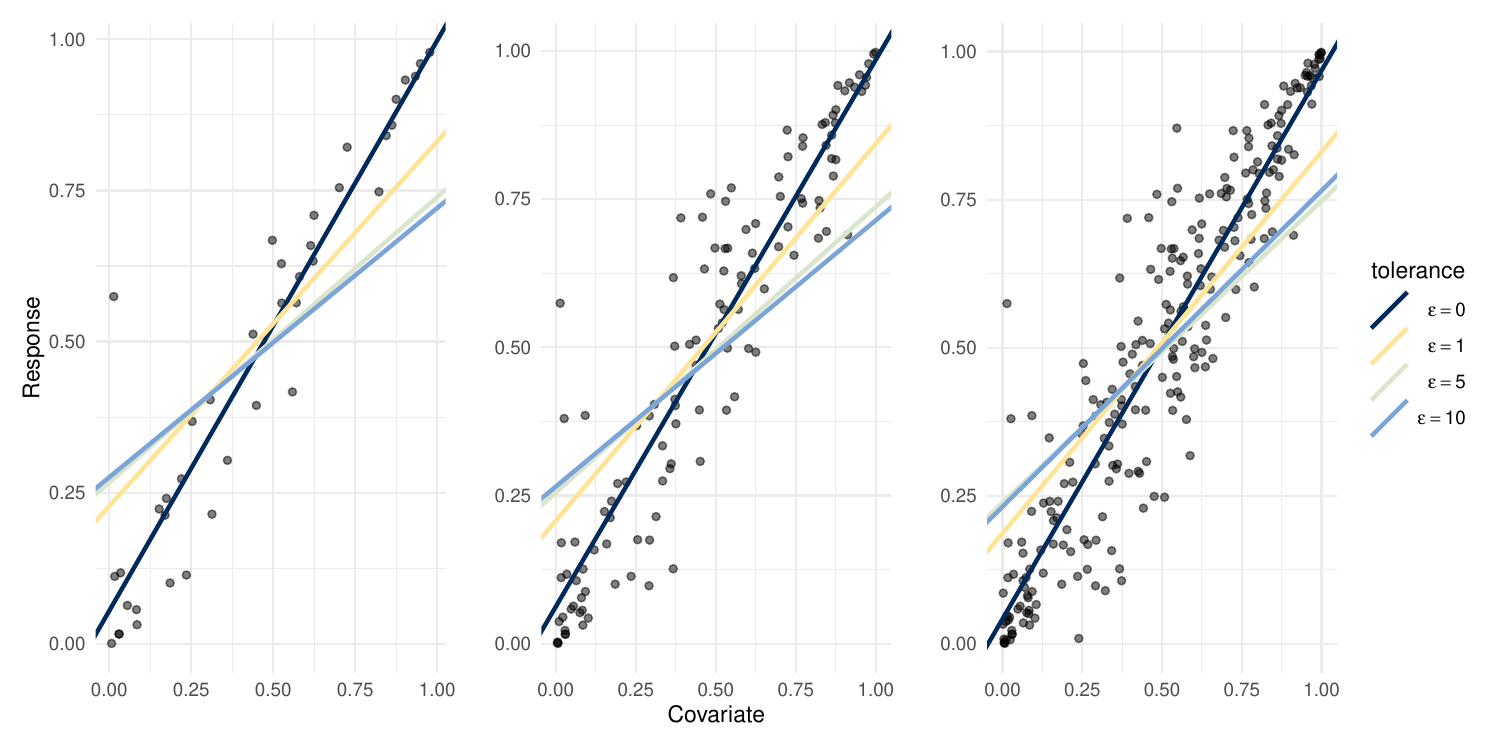}
    \caption{Fitted robust linear regression lines for model A with sample size $n = 40, 80, 120$ (left to right) and for with different $\ep = 0,1,5,10$.}
    \label{fig:rlrn}
\end{figure}
Again, we observe that the smaller  $\ep$, the steeper the slope of the regression lines.

\begin{table}[ht]
    \centering
    \caption{Results of robust linear regression for different dataset sizes $n = 40, 80,$ and $120$. The parameter $\beta_0$ corresponds to the intercept, $\beta_1$ is the slope of the regression line, and MSE the mean squared error.\\}
    \label{tab: ex: RLRn}
    \begin{tabular}{c @{\hspace{2.5em}} c c c l @{\hspace{2.5em}} c c c l @{\hspace{2.5em}} c c c}
   & \multicolumn{3}{c}{$n= 40$} &&  \multicolumn{3}{c}{ $n=80$} && \multicolumn{3}{c}{$n=120$}
    \\
    \cmidrule(lr){2-4} \cmidrule(lr){6-8} \cmidrule(lr){10-12}
    \\
    $\ep$ & $\beta_0$ & $\beta_1$ & MSE & &   $\beta_0$ & $\beta_1$ & MSE & & $\beta_0$  & $\beta_1$ & MSE \\
    \toprule 
    \toprule 
    0 & 0.05 &  0.94  &  0.011 & & 0.06 & 0.92 &  0.012 & &  0.41 &  0.92 &  0.01\\[0.5em]
    1 & 0.22 & 0.60 &  0.023 & & 0.20 & 0.63 &  0.020 & & 0.18 &  0.64 & 0.019\\[0.5em]
    5 & 0.26 & 0.47 &  0.033 & & 0.25 & 0.48 &  0.031 & & 0.23 &  0.50 &  0.027 \\[0.5em]
    10 & 0.27 & 0.44 &  0.033 & & 0.36 & 0.44 &  0.031 & & 0.23 &  0.53  &  0.027\\[0.5em]
    \bottomrule
    \bottomrule
    \end{tabular}
    \end{table}
\Cref{tab: ex: RLRn} reports the regression coefficients and MSEs of the robust linear regressions for the $n = 40, 80, 120$ sample size datasets.  As expected, for linear regression with $\ep = 0$, we have the smallest MSE, as it, by definition, is the solution to minimising the MSE.  The MSE of the regular linear regression remains approximately the same for the three sample sizes, while those of the robust linear regression have a larger deviation. Moreover, for each value of $\ep$, the MSE decreases with increasing sample size.
\end{example}
\vspace{1em}

\subsection{Spanish motor vehicle insurance case study}

We perform expectile regression on a real automotive insurance dataset published in \cite{SeguraLledoPavia2024}. We consider the response to be the premium paid, and the covariates to be the value of the vehicle and the claims made, in tens of thousands of Euro and hundreds of thousand Euro, respectively.  We bootstrap a sample of size $n = 1000$ 30 times, and find the REF expectile regression coefficients $\beta_0, \beta_1, \beta_2$ and baseline expectile regression coefficients, at level $\tau = 0.8$. For the REF expectile regression coefficients, we use the $b$-homogenous scoring function for the expectile, as given in \Cref{prop: expectile}, with $b = 2$.  We compare this to expectile regression as implemented in the \texttt{R} package, \texttt{expectreg} \citep{expectreg}, which corresponds to $\ep = 0$.

\begin{table}[ht]
    \centering
    \caption{Results of bootstrapped robust expectile regression for automotive insurance data. The parameter $\beta_0$ corresponds to the intercept, $\beta_1$ to the claims, $\beta_2$ to the value of the vehicle, and MSE the mean squared error.\\}
    \label{tab: ex: auto}
    \begin{tabular}{ c @{\hspace{2.5em}} c c c l @{\hspace{2.5em}} c c c l @{\hspace{2.5em}} c c c}
    \\
     $\ep$ & $\beta_0$  & $\beta_1$ & $\beta_2$ & MSE\\
    \toprule 
    \toprule
     0 & \phantom{-}0.0187 & 0.0840 & 0.0383 & 0.00032\\[0.5em]
     0.01 & \phantom{-}0.0049 & 0.0455 & 0.1076 & 0.00025\\[0.5em]
     0.05 & -0.0041 & 0.0334 & 0.1570 & 0.00029\\[0.5em]
    \bottomrule
    \bottomrule
    \end{tabular}
    \end{table}

In \Cref{tab: ex: auto}, we report the regression coefficients and MSE for traditional expectile regression ($\ep = 0$), and using the REF methodology with $\ep = 0.01, 0.05$.  We find that using the REF in this circumstance reduced the MSE.  This is in contrast to \Cref{ex:rlr}, where canonical linear regression minimises the MSE by construction.
\begin{figure}[h]
    \centering
    \includegraphics[width=\textwidth]{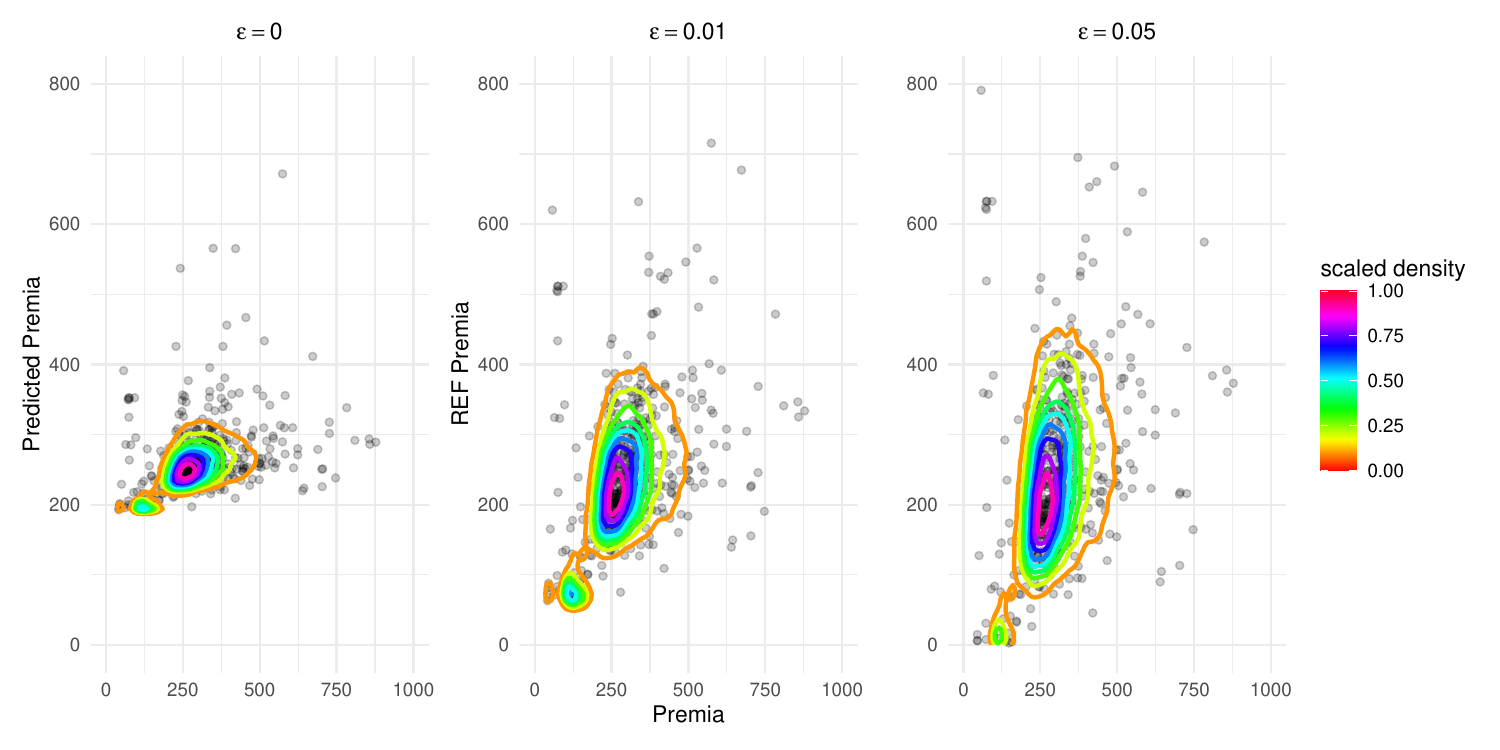}
    \caption{Canonical expectile regression predicted premia and REF predicted premia against actualised premia with $\ep = 0, 0.01, 0.05$ from left to right. Grey points correspond to how real data points are mapped to predicted premia.}
    \label{fig:auto}
\end{figure}

To create \Cref{fig:auto}, we chose $500$ data points are random from the full dataset. The predicted and REF premia are then calculated using the actualised covariates of this sample and plotted against the true premia. Again we interpret increasing $\ep$ as an increase in the joint uncertainty between all covariates and the response. This is supported in \Cref{fig:auto}, where under no uncertainty (left panel), the predicted premia clearly exhibit a dense cluster with a high peak, indicating that the predicted premia are highly concentrated near its median of $259$.  When using the REF, as in the centre and right panels, we see that, while there is a concentration, the density flattens as $\ep$ increases, indicating that there is more variability in the REF premia compared to when $\ep = 0$.

\section{Conclusion}
This paper proposes a new robustification of elicitable functionals, the REF, by incorporating uncertainty prescribed by the KL divergence. Mathematically, the REF is the argmin of an extremal score, where the extremal score is the largest expected score over a set of alternative probability measures that lie within a KL ball. Thus the REF takes the form of a minimax optimisation problem. This methodology is distinct from classical worst-case risk measurement, wherein the risk functional is given as the supremum over all alternative probability measures within the KL ball, of the minimisers of the expected scores.
For the REF, we show that the constraint on the uncertainty region is binding, and characterise conditions for the existence and uniqueness.

Since the REF depends on the choice of scoring function, we explore this choice by using $b$-homogeneous scoring functions, show that these families of scoring functions preserve desirable properties for the REF, and  illustrate them using Murphy diagrams. 

We extend the REF and its representation results to two settings: to $k$-dimensional elicitable functionals, and to functional regression application.  In the $k$-dimensional setting, we consider an application to reinsurance by demonstrating the behaviour of the joint ($\VaR$, $\ES$) REF to a synthetic reinsurance dataset.  In the robust regression setting, we explore the behaviour of the REF under three scenarios: a data contamination problem, a sample size scenario, and a data driven automotive insurance example.

\ACKNOWLEDGMENT{The authors thank Fabio Gomez for stimulating discussions.
SP gratefully acknowledges support from the Natural Sciences and Engineering Research Council of Canada (grants DGECR-2020-00333 and RGPIN-2020-04289) and from the Canadian Statistical Sciences Institute (CANSSI).}

\bibliographystyle{informs2014}
\bibliography{Refs.bib}

\end{document}